\documentclass[showpacs,amsmath,amssymb, nobibnotes, aps, prl,showkeys, floatfix]{revtex4}
\usepackage{graphicx}% Include figure files
\usepackage{dcolumn}% Align table columns on decimal point
\usepackage{bm}% bold math
\usepackage{docs}
\usepackage{amssymb}
\usepackage{amsmath}
\expandafter\ifx\csname package@font\endcsname\relax\else
 \expandafter\expandafter
 \expandafter\usepackage
 \expandafter\expandafter
 \expandafter{\csname package@font\endcsname}%
\fi

\newcommand{\ltsima} {$\; \buildrel < \over \sim \;$}
\newcommand{\gtsima} {$\; \buildrel > \over \sim \;$}
\newcommand{\lta} {\lower.5ex\hbox{\ltsima}}
\newcommand{\gta} {\lower.5ex\hbox{\gtsima}}

\newcommand{\lsim}{\raisebox{-.4ex}{$\stackrel{<}{\scriptstyle \sim}$}}
\newcommand{\gsim}{\raisebox{-.4ex}{$\stackrel{>}{\scriptstyle \sim}$}}

\newcommand{\RNum}[1]{\uppercase\expandafter{\romannumeral #1\relax}}

\begin{document}

\title[Newtonian analogue of Schwarzschild de-Sitter spacetime]{Newtonian analogue of Schwarzschild de-Sitter spacetime: Influence on the local kinematics in galaxies}

\author {Tamal Sarkar \thanks{Email address: ta.sa.nbu@hotmail.com}, Shubhrangshu Ghosh \thanks{Email address: shubhrang.ghosh@gmail.com}, Arunava Bhadra \thanks{Email address: aru\_bhadra@yahoo.com}} 

\affiliation{High Energy $\&$ Cosmic Ray Research Centre, University of North Bengal, Post N.B.U, Siliguri 734013, India. 
 }

\begin{abstract}

The late time accelerated expansion of the Universe demands that even in 
local galactic-scales it is desirable to study astrophysical phenomena, particularly relativistic accretion related phenomena in massive galaxies or in galaxy mergers and the dynamics of the kiloparsecs-scale structure and beyond, in the local-galaxies in Schwarzschild-de Sitter (SDS) background, rather than in Schwarzschild or Newtonian paradigm. Owing to the complex and nonlinear character of the underlying magnetohydrodynamical equations in general relativistic (GR) regime, it is quite useful to have an Newtonian analogous potential containing all the important GR features that allows to treat the problem in Newtonian framework for study of accretion and its related processes. From the principle of conserved Hamiltonian of the test particle motion, here, a three dimensional Newtonian analogous potential has been obtained in spherical geometry corresponding to SDS/Schwarzschild anti-de Sitter (SADS) spacetime, that reproduces almost all of the GR features in its entirety with remarkable accuracy. The derived potential contains an explicit velocity dependent term of the test particle that renders an approximate relativistic modification of Newtonian like potential. The complete orbital dynamics around SDS geometry and the epicyclic frequency corresponding to SDS metric have been extensively studied in the Newtonian framework using the derived potential. Applying the derived analogous potential it is found that the current accepted value of $\Lambda \sim 10^{-56} \rm cm^{-2}$ moderately influences both sonic radius as well as Bondi accretion rate, especially for spherical accretion with smaller values of adiabatic constant and temperature, which might have interesting consequences on the stability of accretion disk in AGNs/radio galaxies. 

\end{abstract}

\maketitle

%\label{firstpage}

\section{Introduction}

The simplest and most attractive general relativistic model that explains the late time accelerated expansion of the Universe [1] is the $\Lambda$CDM model where $\Lambda$ denotes the positive cosmological constant with a value of nearly $10^{-56} \; {\rm cm}^{-2}$ [2] and CDM refers to cold dark matter. The $\Lambda$CDM model is more or less consistent with all the current cosmological observations [3] though the origin of cosmological constant still remains elusive [4] and is one of the most pressing problems in modern physics. 

In the presence of a repulsive cosmological constant (positive) `$\Lambda$' the spacetime geometry exterior to a static spherically symmetric metric is Schwarzschild-de Sitter (SDS), which describes an isolated black hole (BH) in a spatially inflated Universe, rather than Schwarzschild metric.  Therefore, the cosmological constant may affect any local gravitational phenomenon like perihelion shift of the orbits of gravitationally bound systems [5], gravitational bending of light [6], geodetic precession [5] etc., but the general perception is that owing to its tiny value, cosmological constant does not lead to any significant observable effects in a local gravitational phenomenon. In solar system the influence of cosmological constant is known to be maximum in the case of perihelion shift of mercury orbit where the $\Lambda$ contribution is about $10^{-15}$ of the total shift [5]. However, the contribution of repulsive $\Lambda$ could be significant (larger than the second order term) even in a local gravitational phenomenon when kiloparsecs to megaparsecs-scale distances are involved, such as the gravitational bending of light by cluster of galaxies [7]. The study of particle dynamics in SDS spacetime shows a significant contribution of $\Lambda$ when kiloparsecs-scale distance is involved [8,9]. The trajectories of both Small and Large Magellanic Clouds in the gravitational field of the Milky Way are affected significantly ($\sim 10 \%$ level or higher) by $\Lambda$ [10]. Cosmological constant also leads to non-spherical effects with non-negligible contributions on the local dynamics of clusters and superclusters [11]. Some authors claimed [12] that the repulsive $\Lambda$ is responsible for significantly lower value ($\sim 60 \; {\rm Km} \; {\rm s}^{-1}/{\rm Mpc}$) of Hubble parameter in our close neighborhood than its large-scale value ($\sim 70 \; {\rm Km} \; {\rm s}^{-1}/{\rm Mpc}$). The cosmological constant also influences gravitational equilibrium [13].

A plausible local scenario where cosmological constant may contribute significantly is the relativistic accretion phenomena around massive BHs which involve 
distance-scale of the order of hundreds of parsecs or even more. This is because the asymptotic character of the BH spacetime changes significantly in the presence of cosmological constant [7]. Note that the accretion of matter on to BH is believed to act as the engine for the active galactic nuclei (AGNs). A few studies have been carried out so far to investigate the effect of $\Lambda$ in astrophysical jet/accretion flow paradigm [14-19]. It was found that the presence of cosmological constant leads to suppression of black hole evaporation [14]; instead of evaporation black hole will accrete energy. The effect of $\Lambda$ on the dynamics of kiloparsecs to megaparsecs-scale astrophysical jets would be worth intriguing to investigate as suggested by Stuchl\'ik, Sla\'ny and Hled\'ik [15]. Their work indicated that positive $\Lambda$ can have strong collimation effect on jets. Rezzolla et al. [16] have analyzed the effect of $\Lambda$ on the dynamical stability of a geometrically thick accretion disk with constant angular momentum around a SDS BH. Their work showed that $\Lambda$ introduces an outflow of matter through the outer cusp of disk causing a considerable impact on the runway instability. Slobodov et al. [17] demonstrated that characteristic peaks in the iron line profile generated in the BH accretion disk in a SDS background are less noticeable and distinguishable with the increase in the value of $\Lambda$. In an interesting scenario, Karkowski and Malec [18] and Mach, Malec and Karkowski [19] have recently explored the possible dependence of $\Lambda$ on relativistic Bondi-type accretion flow around a nonrotating BH. They defined a quasilocal mass accretion rate and found that $\Lambda$ suppresses the mass accretion rate of the flow and dramatically impacts the transonic nature of the accretion flow.

All the works on the effect of $\Lambda$ on accreting systems are carried out under some restricted conditions/situations. This is because the study of accreting BH systems involves solving general relativistic (GR) hydrodynamic/magnetohydrodynamic (MHD) equations in a strong gravitational field regime. Owing to the complex and nonlinear character of the underlying equations in GR regime, analytical/quasi numerical treatment of the problem is virtually discarded. Even numerical simulation is complicated by several issues such as different characteristic time scales for propagating modes of general relativity and relativistic hydrodynamics. Several early works on these accretion related phenomena were based on pure Newtonian gravity. A few GR effects were incorporated ad hocly. After the seminal work of Paczy\'nski and Witta [20],  most of the authors treated accretion and its related processes around BHs using hydrodynamical/MHD equations in the Newtonian framework by using some PNPs which are essentially modification of Newtonian gravitational potential developed with the objective to reproduce (certain) features of relativistic gravitation. This is to avoid GR gas dynamical equations, which in most occasions become inconceivable in practice in describing a complex physical system like accreting plasma. Consequently, adopting PNPs, one can comprehensively construct more realistic accretion flow models in simple Newtonian paradigm, while the corresponding PNP would capture the essential GR effects in the vicinity of the compact objects. Instead of PNPs, some authors simply use parameterized post Newtonian (PPN) expansion up to a certain order [21]. Since the PPN expansion converges very slowly, the latter option is valid only for orbits at large distances, but not for particle trajectories in the vicinity of the BHs.

Several PNPs exist in the astrophysical literature which are mostly prescribed either in an ad hoc manner or are devised employing certain explicit method [20,22-33], mostly developed for Schwarzschild and Kerr BHs. A PNP corresponding to SDS metric has also been prescribed [34] based on the method in [30-32] for a Keplerian rotation flow which can be more appropriately used to study gas dynamical properties of accretion disk, and thereby concentrating mostly in its use in accretion tori [35], though it has also been recently used to investigate the influence of the repulsive cosmological constant on the kinematics of Small and Large Magellanic cloud [10]. Although notionally PNPs are aimed at mimicking GR geometries, however, they, in general, are devoid of their uniqueness to effectively describe the GR features in its entirety. To replicate GR features, PNPs lay emphasize mostly to reproduce marginally stable and marginally bound orbits. Moreover, in general, they are unable to reproduce observationally verified tests in general relativity like geodetic precession, gravitational bending of light or gravitational time delay.

If the effects of cosmological constant need to be properly revealed in different astrophysical phenomena in local galactic-scales circumventing the complex GR treatment, it is desirable to have a correct Newtonian analogous like potential corresponding to SDS geometry that will reproduce all the salient features of the SDS metric with reasonable accuracy and extensively mimic a wide spectrum of GR behavior. Recently, Tejeda and Rosswog [36] derived a generalized effective potential for a Schwarzschild BH based on a proper axiomatic procedure. Their generalized potential, which has an explicit dependence on radial velocity and orbital angular velocity of test particle, reproduces exactly several relativistic features of corresponding Schwarzschild geometry. Later, the work has been extended for the Kerr BH [37]. In the present work, we would derive a modified Newtonian like potential from the conserved Hamiltonian of the test particle motion [36], which would then be an approximate relativistic potential analogue corresponding to the SDS metric. We would show from a detailed investigation that the derived potential, which depends on the velocity of the test particle, reproduces almost all of the corresponding GR features with remarkable accuracy. It would be worthy enough to explore the feasibility of measurement of cosmological constant in local-scales or in galactic-scales, and further to explore its influence on the dynamics on kiloparsecs-scale structure and beyond in the local-galaxies, or the accretion related scenarios/conditions where cosmological constant may contribute significantly, in which case, the proposed modified Newtonian like potential analogue to SDS geometry considering it as a background metric, would be of remarkable use.

The plan of the paper is the following. In the next section, we would formulate the approximate relativistic potential analogue in the Newtonian framework corresponding to both SDS and Schwarzschild anti-de Sitter (SADS) spacetimes simultaneously, starting from the conserved Hamiltonian of the system in the low energy limit of the test particle motion [36]. Subsequently in \S \RNum{3}, we evaluate the geodesic equations of motion and extensively analyze their solutions corresponding to SDS geometry, with both our analogous potential and general relativity, for the current accepted value of $\Lambda \sim 10^{-56} \rm cm^{-2}$. In \S \RNum{4} we explicitly investigate the influence of repulsive $\Lambda$ on Bondi accretion rate. Finally, we end in \S \RNum{5} with comments and discussion. 

\section{Formulation of a modified Newtonian analogous potential corresponding to SDS/SADS geometry}

For a general class of static spherically symmetric spacetimes of the form (in the standard coordinates system)

\begin{eqnarray}
ds^2 =-f(r)\, c^2 \, dt^2 + \frac{1}{f(r)} \, dr^2 + r^2 d\Omega^2 \, , 
\label{1}
\end{eqnarray}

where, $d\Omega^2 = d\theta^2 +  \sin^2 \theta \, d\phi^2 $ and f(r) is the generic metric function, the Lagrangian density of a particle of mass $m$ is given by

\begin{eqnarray}
2 {\cal L}=-f(r) \, c^2 \,  \left(\frac{dt}{d\tau}\right) ^2 + \frac{1}{f(r)} \, \left(\frac{dr}{d\tau}\right) ^2 +
r^2 \, \left(\frac{d\Omega}{d\tau}\right) ^2\, .  
\label{2}
\end{eqnarray}

From the symmetries, one obtains two constants of motion corresponding to two ignorable coordinates $t$ and $\Omega$ as given by

\begin{eqnarray}
{\cal P}_t =\frac{\partial \cal L}{\partial \tilde{t}}= - c^2 \, f(r)\frac{dt}{d\tau}={\rm constant}=-\epsilon 
\label{3}
\end{eqnarray}               

and 

\begin{eqnarray}
 {\cal P}_{\Omega}=\frac{\partial \cal L}{\partial \tilde{\Omega}} = r^2 \frac{d\Omega}{d\tau}={\rm constant} = \lambda \, ,
\label{4}
\end{eqnarray}

where, $\epsilon$ and $\lambda$ are specific energy and generalized specific angular momentum of the orbiting particle, respectively. Here, $\tilde{t}$ and $\tilde{\phi}$ represent the derivatives of `$t$' and `$\phi$' with respect to proper time $\tau$. It needs to be mentioned that from now onwards, throughout the paper, the terms related to momentum, energy/Hamiltonian, potential and frequency, all of which are in fact their specific quantities, would be addressed without the using of word `specific'. Using Eq. (3) we then can write   

\begin{eqnarray}
\frac{dt}{d\tau} = \frac{\epsilon}{c^2} \frac{1}{f(r)} \, .
\label{5}
\end{eqnarray}

Using $2{\cal L} =g_{\rm {\alpha \beta}} \, p^{\alpha} \, p^{\beta} =-m^2 c^2$ and substituting Eqs. (3) and (4) in Eq. (2), we obtain  

\begin{eqnarray}
\left (\frac{dr}{d\tau} \right)^{2} = \left(\frac{\epsilon^{2}}{c^{2}}- c^{2} \right) 
-c^{2} \, \left[f(r)-1 \right] - f(r) \, \frac{\lambda^2}{r^2}  \, .
\label{6}
\end{eqnarray}

By considering a locally inertial frame for a test particle motion, we write $E_{\rm GN} = (\epsilon^2 - c^4)/{2c^2}$ (`$\rm GN$' symbolizes `GR-Newtonian'). Second term in the above definition of $E_{\rm GN}$ is the rest mass energy of the particle which is subtracted from relativistic energy owing to low energy limit.
 
From Eqs. (5) and (6) and using Eq. (4), we get

\begin{eqnarray}
\frac{dr}{dt} = \frac{c^2}{\epsilon} \, f(r) \sqrt{2 \, E_{\rm GN} 
-c^2 \, \left[f(r)-1 \right] - f(r) \,{\dot \Omega}^2 \frac{r^2}{f(r)} }\, ,
\label{7}
\end{eqnarray}

where, $\dot \Omega$ is the derivative with respect to coordinate time `$t$'. Using the condition for low energy limit $\epsilon/c^2 \sim 1$, as we prefer to, $E_{\rm GN}$ is given by  

\begin{eqnarray}
E_{\rm GN} = \frac{1}{2} \left(\frac{dr}{dt} \right)^2 \frac{1}{f(r)^2} + \frac{r^2 \, {\dot \Omega}^2}{2f(r)} +
\frac{c^2}{2} \, [f(r)-1]  \, .
\label{8}
\end{eqnarray}

In the asymptotic non-relativistic limit $E_{\rm GN}$ reduces to the Newtonian mechanical energy (= Hamiltonian of the motion). The generalized Hamiltonian $E_{\rm GN}$ in the low energy limit should then be equivalent to the Hamiltonian in Newtonian regime. The Hamiltonian in the Newtonian regime with the generalized analogous potential in spherical polar geometry will then be equivalent to $E_{\rm GN}$ in Eq. (8). Thus 

\begin{eqnarray}
E_{\rm GN} \equiv  \frac{1}{2} \left({\dot r}^2 + r^2 {\dot \Omega}^2 \right) 
+ V_{\rm GN} - {\dot r} \frac{\partial V_{\rm GN}}{\partial {\dot r}} 
- {\dot \Omega} \frac{\partial V_{\rm GN}}{\partial {\dot \Omega}} \, , 
\label{9}
\end{eqnarray}

where, $T = 1/2 \, \left({\dot r}^2 + r^2 {\dot \Omega}^2 \right)$ is the non-relativistic kinetic energy of the test particle. $\dot r$ is the derivative with respect to coordinate time `$t$'. $V_{\rm GN}$ is the analogous potential which would then be given by  

\begin{eqnarray}
V_{\rm GN}  = \frac{c^2 [f(r)-1]}{2}- \frac{[1-f(r)}{2f(r)} 
\left[ \frac{1+f(r)}{f(r)} \dot r^{2} + r^{2} \, {\dot \Omega}^2 \right] \, .
\label{10}
\end{eqnarray}

$V_{\rm GN}$, thus, is the generalized three dimensional potential in spherical geometry in Newtonian analogue, corresponding to any generalized static GR metric given in Eq. (1), with test particle motion in the low energy limit. Note that, the first term on the right hand side of the potential contains the explicit information of the source. For a purely spherically symmetric gravitational mass with zero charge and without any external effects, the classical Newtonian gravitational potential $-GM/r$ will be recovered from this term. The second term is the explicit velocity dependent and contains the information of the test particle 
motion, thus contributing to the modification of Newtonian gravity. 

For SDS/SADS metric the metric function $f(r)= 1- \frac{2r_s}{r} - \frac{\Lambda r^2}{3}$, where $\Lambda$ is the cosmological constant. $\Lambda > 0$ represents the SDS metric with spatially inflated Universe, where $\Lambda < 0$ represents a SADS metric corresponding to negative vacuum energy density. $r_s = GM/c^2$. From Eq. (10) we then obtain the three dimensional generalized Newtonian analogous potential in spherical geometry, corresponding to SDS/SADS geometry in the low energy limit, which is given by   

\begin{eqnarray}
V_{\rm DS/ADS}  = -\left( \frac{GM}{r} + \frac{\Lambda \, c^{2} \, r^{2}}{6}  \right)
- \left(\frac{2r_s + \frac{\Lambda \, r^3}{3}}{r-2r_s - \frac{\Lambda \, r^3}{3}} \right) \left( \frac{r-r_s - \frac{\Lambda \, r^3}{6}}{r-2r_s - \frac{\Lambda \, r^3}{3}} \dot r^2 + \frac{r^2 \, {\dot \Omega}^2}{2} \right) \, , \nonumber \\
\label{11}
\end{eqnarray}

where, subscript `DS/ADS' denotes Schwarzschild de-Sitter/anti-de Sitter. With $\Lambda = 0$, the above potential reduces to the potential corresponding to the simplest static Schwarzschild geometry, given in [36]. $M \equiv M_{\rm BH}$ is the mass of the BH/central object. The denominator in the second term of the potential contains the exact metric function $f(r)$ of SDS/SADS geometry, and hence the potential $V_{\rm DS/ADS}$ would reproduce the exact location of the event horizon and cosmological horizon and other horizon properties, as that in full general relativity. Introducing a dimensional parameter or cosmological parameter ${\zeta} = {\Lambda r^2_s}/3$, the vanishing cubic polynomial $f(r)$ with repulsive cosmological constant ($\Lambda > 0$) would give two real positive roots representing the locations of two horizons, namely the BH horizon and the cosmological horizon. The locations of these two horizons are then given by $r_H = \frac{2}{\sqrt{3 {\zeta}}} \, \cos \left[\frac{\pi}{3} + \frac{\cos^{-1} \left(3 \, \sqrt{3 {\zeta}} \right) }{3} \right]$ and $r_{\rm CM} = \frac{2}{\sqrt{3 {\zeta}}} \, \cos \left[\frac{\pi}{3} - \frac{\cos^{-1} \left(3 \, \sqrt{3 {\zeta}} \right) }{3} \right]$, respectively.

In the next section, we will analyze different aspects of particle dynamics in the gravitational field of SDS geometry in the modified Newtonian framework.

\section{Orbital dynamics around SDS spacetime}

In the Newtonian framework, the Lagrangian of a particle in the presence of the SDS analogous potential $V_{\rm DS}$ per unit mass is given by, 

\begin{eqnarray}
{\cal L}_{\rm DS} = \frac{1}{2} \left[\frac{r^2 \, {\dot r}^2}{\left(r-2r_s - \frac{\Lambda \, r^3}{3}\right)^2} + \frac{r^3 \, {\dot \Omega}^2}{r-2r_s - \frac{\Lambda \, r^3}{3}} \right] \, + \, \frac{GM}{r} + \frac{\Lambda \, c^2 r^2}{6} \, , 
\label{12}
\end{eqnarray}

where, ${\dot \Omega}^2= {\dot \theta}^2 + \sin^2 \theta \, {\dot \phi}^2$. Here overdots denote the derivative with respect to coordinate time `$t$'. We then compute the conserved angular momentum and Hamiltonian using $V_{\rm DS}$, given by 

\begin{eqnarray}
\lambda_{\rm DS} = \frac{r^3 \, \dot \Omega}{r-2r_s - \frac{\Lambda \, r^3}{3}}  
\label{13}
\end{eqnarray}

and 

\begin{eqnarray}
 E_{\rm DS} =  \frac{1}{2} \left[\frac{r^2 \, {\dot r}^2}{\left(r-2r_s - \frac{\Lambda \, r^3}{3}\right)^2} + \frac{r^3 \, {\dot \Omega}^2}{r-2r_s - \frac{\Lambda \, r^3}{3}} \right] \, - \, \frac{GM}{r} - \frac{\Lambda \, c^2 r^2}{6} \, ,  
\label{14}
\end{eqnarray}

respectively. Using Eqs. (13) and (14), we obtain $\dot r$ that uniquely describes the test particle motion, which is given by 

\begin{eqnarray}
\frac{dr}{dt} =  \frac{r-2r_s - \frac{\Lambda \, r^3}{3}}{r}
 \sqrt{2 E_{\rm DS}  + \frac{2GM}{r} + \frac{\Lambda \, c^{2} r^2}{3} - 
\left(r-2r_s - \frac{\Lambda \, r^3}{3} \right) \frac{\lambda^2_{\rm DS}}{r^3} }\, , \nonumber \\
\label{15}
\end{eqnarray}

which is exactly equivalent to $\dot r$ in general relativity in low energy limit. Replacing $\dot \Omega$ and $\dot r$ in Eq. (11) using Eqs. (13) and (15) respectively,  SDS analogous potential can be written in terms of conserved Hamiltonian $E_{\rm DS}$ and angular momentum $\lambda_{\rm DS}$, given by 

\begin{eqnarray}
V_{\rm DS}  = -\left( \frac{GM}{r} + \frac{\Lambda \, c^{2} \, r^{2}}{6} \right)
- \left(2r_s + \frac{\Lambda \, r^3}{3}\right) \left[\left(r-2r_s - \frac{\Lambda \, r^3}{3}\right) \frac{\lambda^2_{\rm DS/ADS}}{r^4} \left(\frac{1}{2} - \frac{r - r_s - \frac{\Lambda \, r^3}{6}}{r} \right)\right]  \nonumber \\
- \left(2r_s + \frac{\Lambda \, r^3}{3}\right) \left[\frac{1}{r^2} \left(r - r_s - \frac{\Lambda \, r^3}{6}\right) \left(2 E_{\rm DS/ADS}  + \frac{2GM}{r} + \frac{\Lambda \, c^{2} r^2}{3} \right)  \right]
\label{16}
\end{eqnarray}

In Fig. 1a we show the variation of $V_{\rm DS}$ in the form given in Eq. (16) with the radial distance $r$, corresponding to $\Lambda > 0$ with $\Lambda r^2_s = 1 \times 10^{-27}$ in the low energy limit of test particle motion. The stated value of $\Lambda r^2_s$ corresponds to $\Lambda = 10^{-56} {\rm cm^{-2}}$ for $M_{\rm BH} \sim 10^9 \, M_{\odot}$. Unless stated explicitly, we shall use such a combination of $\Lambda$ and $M_{\rm BH}$ throughout the paper. The profile of $V_{\rm DS}$ clearly shows both the BH event horizon as well as the cosmological horizon. With $\Lambda r^2_s \sim 1 \times 10^{-27}$, cosmological horizon extends up to $\sim 5.5 \times 10^{13} \, r_s$. For a BH of $\sim 10^9 M_{\odot}$, it gives a radius of $\sim 5.5 \times 10^3 \, {\rm megaparsec}$. Figure 1b resembles that of 1a but for semi-relativistic test particle energy ($E_{\rm DS} \sim 0.5$). Although the locations of BH event horizon as well as cosmological horizon remain unaltered with the increase in test particle energy, nevertheless, there is a noticeable change in the magnitude of $V_{\rm DS}$ at both horizon radii. It is found that 
$\lambda_{\rm DS}$ has no effect on the nature of potential just beyond $\sim 100 \, r_s$. These figures indicate that with the increase in the value $\lambda_{\rm DS}$, inner horizon shifts to larger radii. 

Figures 1c and 1d show the variation of effective potential $V^{\rm Eff}_{\rm DS} = V_{\rm DS} +  \frac{\lambda^2_{\rm DS}} {2 r^4} \left(r-2r_s - \frac{\Lambda \, r^3}{3}\right)^2$ for the same value of $\Lambda$ as in figures 1a,b, corresponding to SDS spacetime. The nature of the profiles resemble that in figures 1a,b, however $V^{\rm Eff}_{\rm DS}$ attains a higher peak as compared to $V_{\rm DS}$ in the vicinity of inner horizon for values of $\lambda_{\rm DS} \, \gsim \, 3.5 $. The figure 1c shows the variation of `GR effective potential' [see Eq. (30)] for the same values of angular momentum.  

\begin{figure}
\centering
\includegraphics[width=1.0\columnwidth]{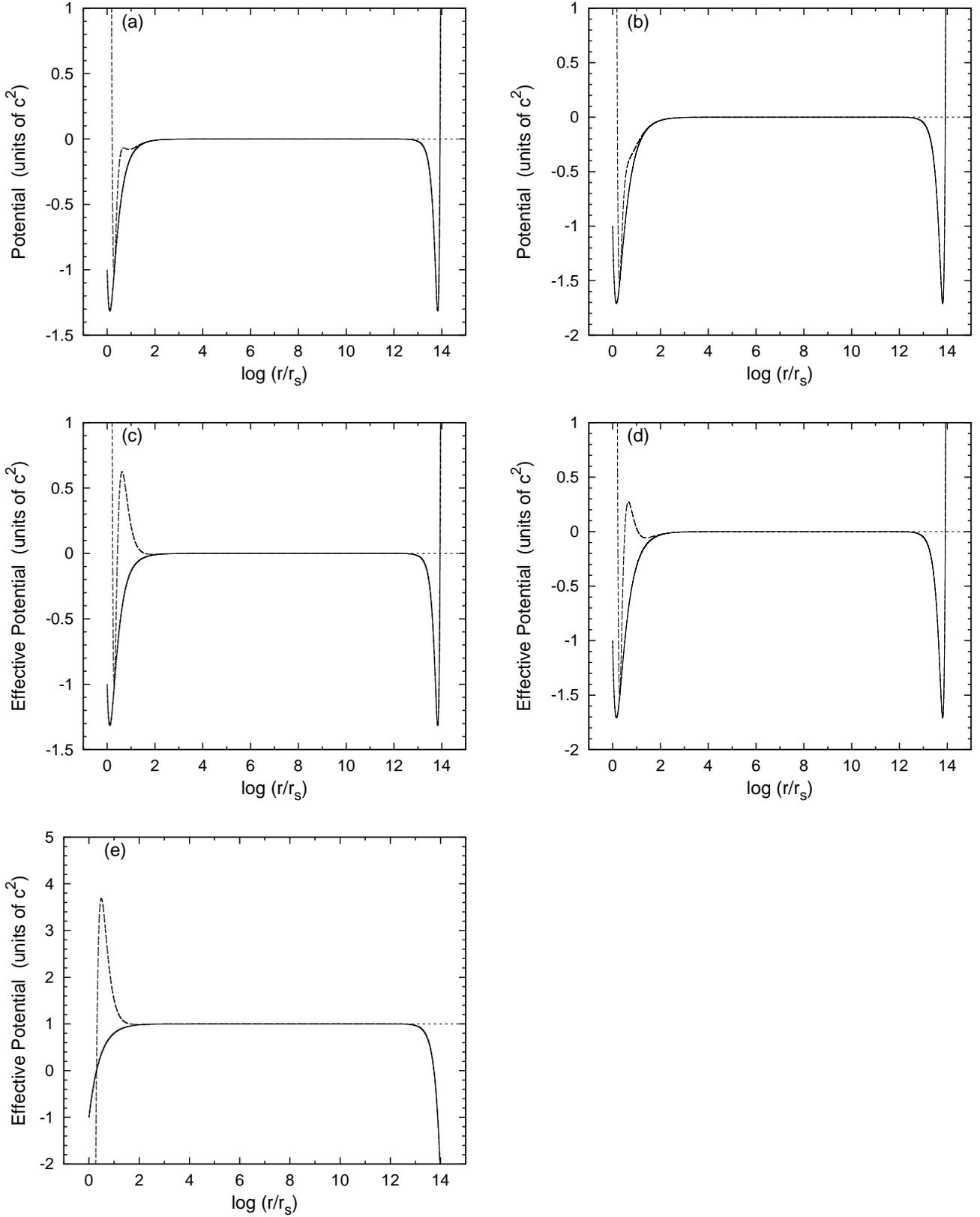}
\caption{Variation of potential $V_{\rm DS}$ [Eq. (16)] with radial distance $r$. The solid and long-dashed lines in figure 1a correspond to $V_{\rm DS}$ in low energy limit $E_{\rm DS} \sim 0$ for SDS geometry with $\Lambda  r^2_s= 1 \times 10^{-27} $ with $\lambda_{\rm DS} = (0, 9.5)$, respectively. The short-dashed curve represents Schwarzschild spacetime ($\Lambda = 0$). Figure 1b is for semi-relativistic energy $E_{\rm DS} \sim 0.5$, otherwise resemble 1a. Figures 1c and 1d correspond to effective potential $V^{\rm Eff}_{\rm DS}$, with other  parameters same to those in figures 1a and 1b, respectively. Figure 1e corresponds to `GR effective potential' for SDS geometry with the same physical parameters as those in figures 1c,d, but independent of conserved energy. $E_{\rm DS}$ and $\lambda_{\rm DS}$ are expressed respectively in units of $c^2$ and $GM/c$.    
 } 
\label{Fig1}
\end{figure}

In Fig. 2a, we depict the difference in the magnitude between $V_{\rm DS}$ in Eq. (16) and the analogous potential for Schwarzschild metric in [36] $\left(V_{\rm SW}\right)$, with radial distance $r$. For semi-relativistic energy of test particle, the magnitude of $V_{\rm DS}$ may even exceed the corresponding Schwarzschild value by $\gsim \, 1.6 c^2$ at the outer radii near the cosmological horizon. In Fig. 2b, we show the relative deviation (in percentage) between the $V_{\rm DS}$ and $V_{\rm SW}$, as a function of radial distance $r$. For any arbritary physical quantity $\mathcal F$, the relative deviation $(\xi_{\imath})$ is defined as $\xi_{\imath} = 2 \lvert \frac{{\mathcal F}_{\rm DS} -  {\mathcal F}_{\rm SW} }{ {\mathcal F}_{\rm DS} + {\mathcal F}_{\rm SW} } \rvert$ which essentially implies deviation between the 
analogous potential for SDS and Schwarzschild geometry relative to their average value. The subscript `SW' represents corresponding quantities in `Schwarzschild' geometry. 

It is found from the Fig. 2b that $\xi_{\imath}$ sharply increases with $r$, especially, at the outer radii. At smaller distances $\xi_{\imath}$ is very small,  and detecting such small deviations experimentally does not seem to be possible at present or in near future. At what distance $\xi_{\imath}$ can be said as non-negligible depends entirely on experimental accuracy. The solar system experiments presently are capable to detect any deviation from general relativity at $10^{-5}$ level. If we demand a similar experimental accuracy, we can then say that when $\xi_{\imath}$ attains a value of $\sim 10^{-4}$ (or equivalently $\sim 0.01 \%)$ the relative deviation may be considered to be non-negligible with the corresponding distance $r \sim 8.4 \times 10^7 \, r_s$. With this consideration, this then implies that beyond such radius the influence of $\Lambda$ cannot be neglected. At $r \, \gsim \, 8.5 \times 10^8 \, r_s$, the relative deviation becomes substantial with $\left(\xi_{\imath} \, \gsim \, 10 \% \right)$. Figure 2b also shows that $\xi_{\imath}$ increases rapidly up to a certain radius $r \sim 9 \times 10^9 \, r_s$, beyond which it mostly remains constant. Around this particular radius, the $\Lambda$ effects become 
too dominating, which renders $V_{\rm SW}$ to become negligible in compare to that of $V_{\rm DS}$. We describe this radius $r \sim 9 \times 10^9 \, r_s$ as some upper bound $\left(\equiv x_{\rm max} \right)$, where $\xi_{\imath}$ attains a values of $\sim 200 \%$. For $M_{\rm BH} \sim 10^9 \, M_{\odot}$, in accordance with the BH mass in many AGNs/quasars or in massive galaxies, the lower bound in $r \left(\equiv x_{\rm min} \right)$, gives a radius of $\sim 8$ kiloparsec, whereas the upper bound $x_{\rm max}$ gives a radius of $\sim 900$ kiloparsec. Thus, the region between $x_{\rm min}$ and $x_{\rm max}$ or a region approximately existing between few kiloparsecs to a few $100$ kiloparsecs would be strongly affected by cosmological constant $\Lambda$, thereby, directly influencing the kiloparsecs-scale structure in massive galaxies, in the local observable Universe. With a similar BH mass, the region beyond $80$ kiloparsec would have most significant $\Lambda$ effects, where $\xi_{\imath}$ would be greater than $\sim 10 \%$. 

\begin{figure}
\centering
\includegraphics[width=1.0\columnwidth]{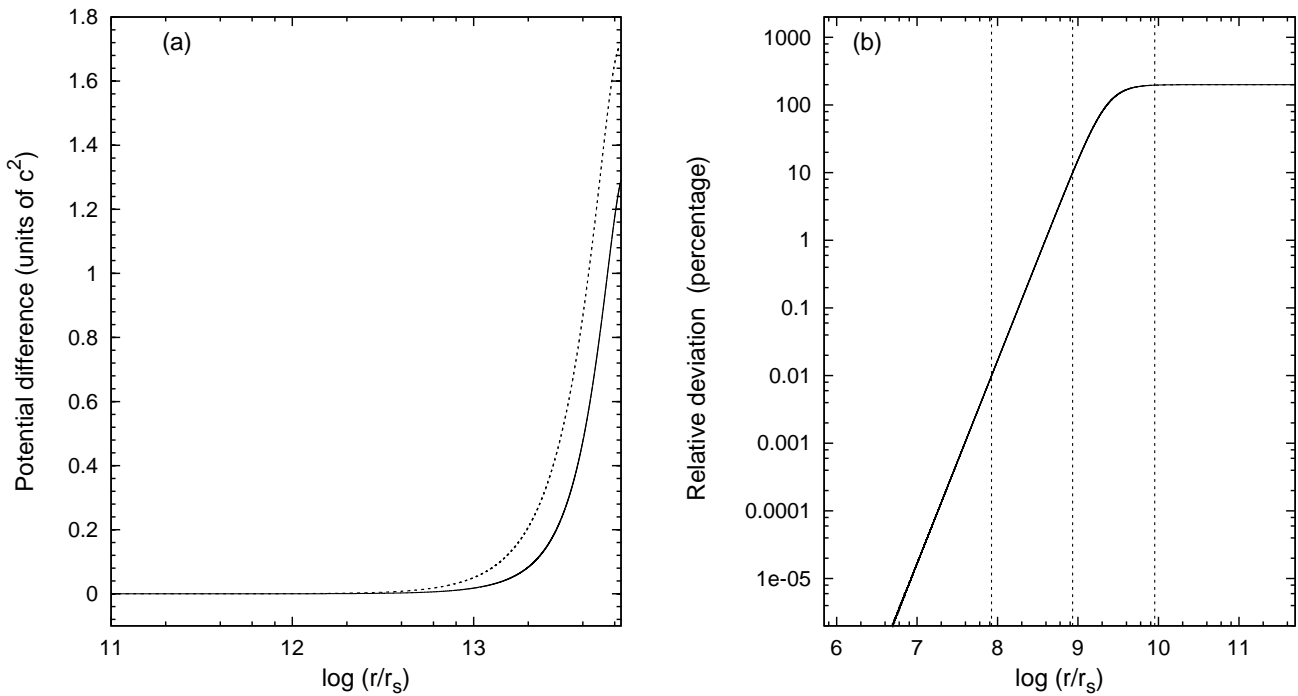}
\caption{Variation of the difference between $V_{\rm DS}$ ($\Lambda r^2_s = 1 \times 10^{-27}$) and $V_{\rm SW}$ with $r$. Solid and short-dashed curves in Fig. 2a show the difference between $V_{\rm DS}$ and $V_{\rm SW}$, corresponding to $E_{\rm DS} \sim 0$ and $E_{\rm DS} \sim 0.5$, respectively. Solid curve in Fig. 2b shows the variation of the relative deviation (in percentage) $\xi_{\imath}$ with $r$, for both $E_{\rm DS} \sim 0$ and $E{\rm DS} \sim 0.5$ (which coincide). The two extreme vertical dashed lines, denoted as $x_{\rm min}$ and $x_{\rm max}$ (in units of $r_s$, correspond to the value of $\xi_{\imath}$ $\sim 0.01 \%$ and 
$ \sim 200 \%$, respectively). The vertical dashed line in the middle represents $r$ at which $\xi_{\imath} \sim 10 \%$. $E_{\rm DS}$ is expressed in units of $c^2$. Note that $\lambda_{\rm DS}$ has no effect on the nature of potentials when $r > 100 \, r_s$.
 } 
\label{Fig2}
\end{figure}

Next we obtain the equation of the orbital trajectory using Eqs. (13) and (15), as given by

\begin{eqnarray}
\left(\frac{dr}{d\Omega}\right)^2 = \frac{r^4}{\lambda^2_{\rm DS/ADS}} \left[2 E_{\rm DS/ADS}  + \frac{2GM}{r} + \frac{\Lambda \, c^{2} r^2}{3}  
- \left(r-2r_s - \frac{\Lambda \, r^3}{3} \right) \frac{\lambda^2_{\rm DS/ADS}}{r^3} \right] \, . \nonumber \\
\label{17}
\end{eqnarray}

Although Eq. (17) is derived with the condition of low energy limit, yet it is exactly the same as that in full general relativity which one can easily obtain using Eqs. (4) and (7) with relevant $f(r)$. To furnish a complete behavior of the particle dynamics in SDS background, we obtain the equations of motion of the test particle in the presence of $V_{\rm DS}$ from the Euler-Lagrange equations in spherical geometry which are given by 

\begin{eqnarray}
\ddot r = \left( -\frac{GM}{r^2} + \frac{\Lambda \, c^2 r}{3} \right) 
\left(\frac{r-2r_s - \frac{\Lambda \, r^3}{3}}{r} \right)^{2} 
\, + \, \frac{2 \left(r_s - \frac{\Lambda \, r^3}{3} \right)}{r \left(r-2r_s - \frac{\Lambda \, r^3}{3}  \right)} \, {\dot r}^2 + \left(r-3 r_s \right) 
\left({\dot \theta}^2 + \sin^2 \theta \, {\dot \phi}^2 \right) \, ,
\label{18}
\end{eqnarray}

\begin{eqnarray}
\ddot \phi = -\frac{2 \, \dot r \dot \phi}{r} \left(\frac{r-3r_s}{r - 2 r_s - \frac{\Lambda \, r^3}{3}} \right) \, - \, 2 \, \cot \theta \, \dot \phi \, \dot \theta 
\label{19}
\end{eqnarray}

and 

\begin{eqnarray}
\ddot \theta = -\frac{2 \, \dot r \dot \theta}{r} \left(\frac{r-3r_s}{r - 2 r_s - \frac{\Lambda \, r^3}{3}} \right) \, + \, \sin \theta \, \cos \theta \, 
{\dot \phi}^2 \, ,
\label{20}
\end{eqnarray}

respectively. $\ddot \phi$ and $\ddot \theta$ equations are exactly the same to that in general relativity, whereas $\ddot r$ equation in (18) corresponds to that in general relativity in the low energy limit. The corresponding $\ddot r$ equation in general relativity is given by 

\begin{eqnarray}
\ddot r = \left( -\frac{GM}{r^2} + \frac{\Lambda \, c^2 r}{3} \right) 
\left(\frac{r-2r_s - \frac{\Lambda \, r^3}{3}}{r} \right)^{2} \, 
\frac{c^4}{\epsilon^2} 
\, + \, \frac{2 \left(r_s - \frac{\Lambda \, r^3}{3} \right)}{r \left(r-2r_s - \frac{\Lambda \, r^3}{3}  \right)} \, {\dot r}^2 + \left(r-3 r_s \right) 
\left({\dot \theta}^2 + \sin^2 \theta \, {\dot \phi}^2 \right) \, .
\label{21}
\end{eqnarray}

It needs to be mentioned that previously a PNP has been prescribed in [34] corresponding to SDS geometry based on a method adopted by [30-32], which has been formulated by considering a Keplerian rotation profile of the test particle motion. The form of the PNP is given by 

\begin{eqnarray}
\Phi (r) = \frac{r^3 \frac{\Lambda}{3} - 3 r \left(\frac{\Lambda}{3}\right)^{1/3} + 2}{2 \left[1-3\left(\frac{\Lambda}{3}\right)^{1/3} \right] \left(2-r+ r^3 \frac{\Lambda}{3}\right) } \, ,
\label{22}
\end{eqnarray}

which is derived on the premise that $\Phi (r) = 0$ at the static radius, preserving the analogy that the gravitational potential tends to zero in asymptotically flat spacetime. This potential does not have any dependence on test particle velocity. With $\Lambda = 0$, the PNP in Eq. (22) reduces to that of Paczy\'nski-Witta potential corresponding to Schwarzschild geometry. The corresponding GR behavior is mimicked through this PNP by intending only to reproduce the marginally stable and bound orbits for circular orbital trajectory. This is in sharp contrast with the velocity dependent potential $V_{\rm DS}$, which, a priori, focused on to replicate the general relativity by resembling the geodesic equations of motion. This ensures that most of the GR features could be reproduced accurately. The equation of the orbital trajectory and the equations of motion for the corresponding PNP in Eq. (22) in spherical geometry are then given by 

\begin{eqnarray}
\left(\frac{dr}{d\Omega}\right)^2 = \frac{r^4}{\lambda^2} \left[2 E  + \frac{r^3 \frac{\Lambda}{3} - 3 r \left(\frac{\Lambda}{3}\right)^{1/3} + 2}{2 \left[1-3\left(\frac{\Lambda}{3}\right)^{1/3} \right]\left(2-r+ r^3 \frac{\Lambda}{3}\right)}  
-  \frac{\lambda^2}{r^3} \right] \, , 
\label{23}
\end{eqnarray}

\begin{eqnarray}
\ddot r = - \frac{\left(1- r^3 \frac{\Lambda}{3}\right)}{\left(2-r+ r^3 \frac{\Lambda}{3}\right)^2 } +
r \left({\dot \theta}^2 + \sin^2 \theta \, {\dot \phi}^2 \right) \, ,
\label{24}
\end{eqnarray}

\begin{eqnarray}
\ddot \phi = -\frac{2 \, \dot r \dot \phi}{r} - \, 2 \, \cot \theta \, \dot \phi \, \dot \theta 
\label{25}
\end{eqnarray}

and 

\begin{eqnarray}
\ddot \theta = -\frac{2 \, \dot r \dot \theta}{r}  + \, \sin \theta \, \cos \theta \, 
{\dot \phi}^2 \, ,
\label{26}
\end{eqnarray}

It is seen that the trajectory equation as well as the equations of motion corresponding to the PNP in Eq. (22) do not at all resemble the equivalent relations in general relativity and hence this PNP cannot reproduce the features of SDS spacetime accurately. Notwithstanding, in the next few sections we would compare the dynamical behavior of the test particle motion obtained with $V_{\rm DS}$ and the PNP in Eq. (22) as well as with full general relativity, over the entire spatial regime relevant for SDS background. 

\subsection{Particle dynamics along circular orbit}

In order to compare the behavior of the particle motion in presence of $V_{\rm DS}$ and those in general relativity, we compute the dynamical variables for the simplest circular orbit trajectory. With the conditions for the circular orbits $\dot r = 0$ and $\ddot r=0$, we obtain corresponding angular momentum $\lambda^C_{\rm DS}$, Hamiltonian $E^C_{\rm DS}$ and the orbital angular velocity $\dot \Omega^{C}_{\rm DS}$ with $V_{\rm DS}$ using Eqs. (13), (15) and (18), given by 

\begin{eqnarray}
\lambda^C_{\rm DS} = r \sqrt{\frac{GM - \frac{\Lambda \, c^{2} r^3}{3}}{r-3r_s} } \, ,
\label{27}
\end{eqnarray}

\begin{eqnarray}
E^C_{\rm DS} = - \frac{GM}{2r} \left( \frac{r-4r_s}{r-3r_s} \right) + 
\left[\frac{\Lambda \, c^2 r^3}{6(r-3r_s)} \right] \, \left(\frac{\Lambda r^2}{3} +  \frac{4r_s}{r} - 2 \right) \,
\label{28}
\end{eqnarray}

and 

\begin{eqnarray}
\dot \Omega^{C}_{\rm DS} = \frac{r - 2r_s - \frac{\Lambda \, r^3}{3} }{r^2} 
\sqrt{\frac{GM - \frac{\Lambda \, c^2 r^3}{3}}{r-3r_s} } \, ,  
\label{29}
\end{eqnarray}

respectively. With $\Lambda = 0$, all the above equations get reduced to that obtained in Schwarzschild geometry. To compute these variables in exact SDS geometry we use the corresponding `GR effective potential', given by 

\begin{eqnarray} 
V^{\rm GR}_{\rm eff} \, (r) = \left(1 - \frac{2 r_s}{r} - \frac{\Lambda \, r^2}{3}\right) \left(c^2 + \frac{\lambda^2}{r^2} \right) \, .
\label{30}
\end{eqnarray}

As usual, circular orbits occur in general relativity when ${dr}/{d\tau} = 0 $ and ${\partial V^{\rm GR}_{\rm eff}}/{\partial r} = 0$. We obtain energy $\epsilon$ for particle motion in circular orbit, given by 

\begin{eqnarray} 
\frac{\epsilon}{c^2} = {\frac{\left(r - 2r_s - \frac{\Lambda \, r^3}{3}\right)}{\sqrt{r (r-3r_s)}}}  \, .
\label{31}
\end{eqnarray}

Angular momentum ($\lambda^{C}$) and the equivalent Hamiltonian $E^C$ $\left[= \left(\epsilon^2 - c^4 \right)/{2c^2} \right]$ for circular orbits in general relativity then exactly resemble the corresponding values obtained with $V_{\rm DS}$, given by Eqs. (27) and (28) respectively. The orbital angular velocity in general relativity is then given by 

\begin{eqnarray}
\dot \Omega^{C} = \sqrt{\frac{GM}{r^3} - \frac{\Lambda \, c^2}{3} } \, , 
\label{32}
\end{eqnarray}

whose analytical expression is not exactly equivalent to $\dot \Omega^{C}_{\rm DS}$ in Eq. (29). Note that the circular orbit corresponding to SDS metric exists down to $3 \, r_s$ in similarity to that in Schwarzschild geometry which represents the null hypersurface. The photon orbit is thus independent of the cosmological constant. For PNP in Eq. (22), the corresponding expressions of $\lambda^C$, $E^C$ and $\dot \Omega^C$ are given in [34]. The said PNP does not reproduce the correct photon orbit.

In Fig. 3 we show the appropriate comparison of the nature of potential $V_{\rm DS}$ in Eq. (11) and that of PNP in Eq. (22), corresponding to test particle motion in circular orbit. In the very inner and outer regions near to both the BH and cosmological horizons, the behavior of these potentials differ significantly (figures 3c,d). One of the major distinctive features of the PNP in Eq. (22) is that it becomes zero at the static radius $\left(\sim 1.4 \times 10^9 \, r_s \right)$, whereas $V_{\rm DS}$ attains a value of $\, \sim - 10^{-9} \, c^2$ (Fig. 3b). In the intermediate region, the nature of the potentials remain mostly similar. It is seen that at $r \, \gsim \, 10^8 \, r_s$, where the effect of $\Lambda$ is prominent, the PNP in Eq. (22) differs significantly as compared to $V_{\rm DS}$. This radius approximately resembles $x_{\rm min}$ in Fig. 2b. 

\begin{figure}
\centering
\includegraphics[width=1.0\columnwidth]{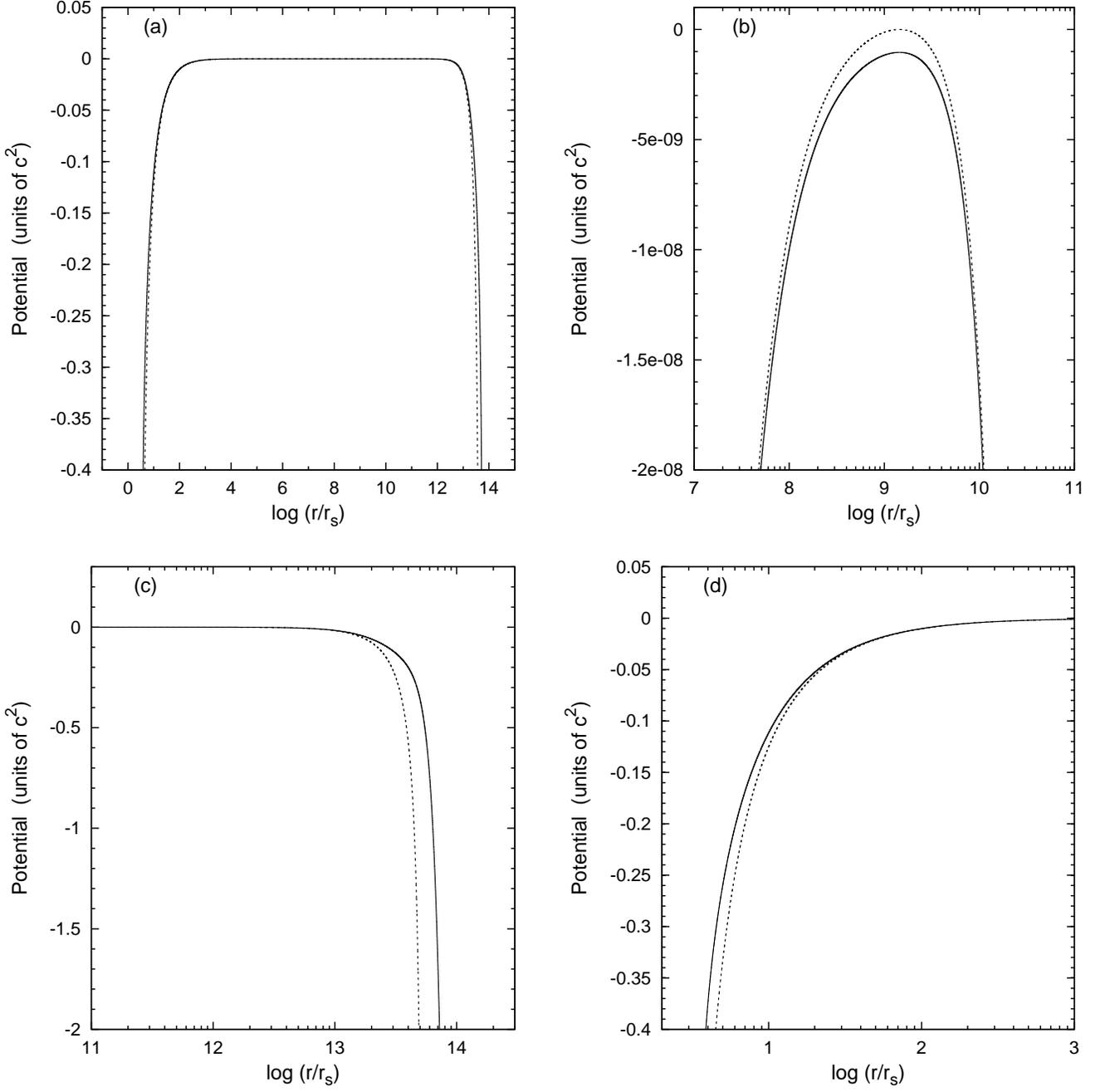}
\caption{Variation of potential $V_{\rm DS}$ and the PNP in Eq. (22) with $r$ corresponding to circular motion trajectory for $\Lambda r^2_s= 1 \times 10^{-27}$. The solid and short-dashed lines correspond to $V_{\rm DS}$ in Eq. (11) and PNP in Eq. (22), respectively. Figure 3a covers the entire distance, figures 3b,c focus only on the outer radii; near to the static radius and the cosmological horizon, respectively, whereas figure 3d focuses on the inner radii near the BH event horizon. 
 }
\label{Fig3}
\end{figure} 

In Fig. 4, we furnish the variation of $\lambda^C_{\rm DS}$ corresponding to $V_{\rm DS}$ which exactly coincides with that in general relativity, for the entire spatial regime. In Fig. 4b, we compare the nature of $\lambda^C_{\rm DS}$ with that corresponding to the PNP in Eq. (22) at the outer radii. The figures show that for SDS spacetime, there is a clear static radius at the outer radii where angular momentum abruptly falls to zero value. For BH of $\sim 10^9 M_{\odot}$, with $\Lambda = 10^{-56} {\rm cm^{-2}}$, the static radius will be located at $\sim 140 \, {\rm kiloparsec}$. Figure 4b shows that the angular momentum for circular orbit trajectory corresponding to $V_{\rm DS}$ as well as for the PNP in Eq. (22) behaves 
quite similarly at the outer radii, and coincides near the static radius. However, due to the profound effect of $\Lambda$ on the outer radii, the nature of 
$\lambda^C_{\rm DS}$ at the outer radii differs significantly from that corresponding to $V_{\rm SW}$, especially beyond $4 \times 10^8 \, r_s$. In the inner radii, where the effect of $\Lambda$ is negligible, $\lambda^C_{\rm DS}$ show appreciable deviation from the angular momentum profile corresponding to the PNP in Eq. (22). For a comparison 
of the profiles of different PNPs in the inner radii, the readers should see [36]. Resembling Fig. 2b, in Fig. 5, we show the relative deviation $\left(\xi_{\imath}\right)$ in angular momentum as a function of radial distance $r$, showing approximately similar behavior as that in Fig. 2b. For this profile (Fig. 5), we found that the value of $x_{\rm min} \sim 8.4 \times 10^7 $ and the value of $x_{\rm max} \sim 1.4 \times 10^9 $, corresponding to the value of $\xi_{\imath}$ $\sim 0.01\%$ and $\sim 200 \%$, respectively. At $x_{\rm max}$, which is also the static radius, the curve gets truncated. At $r \, \gsim  \, 8 \times 10^8 \, r_s$, where the effect of $\Lambda$ is significant, large relative deviation ($\gsim \, 10 \%$) in angular momentum profile is noticed. 

\begin{figure}
\centering
\includegraphics[width=1.0\columnwidth]{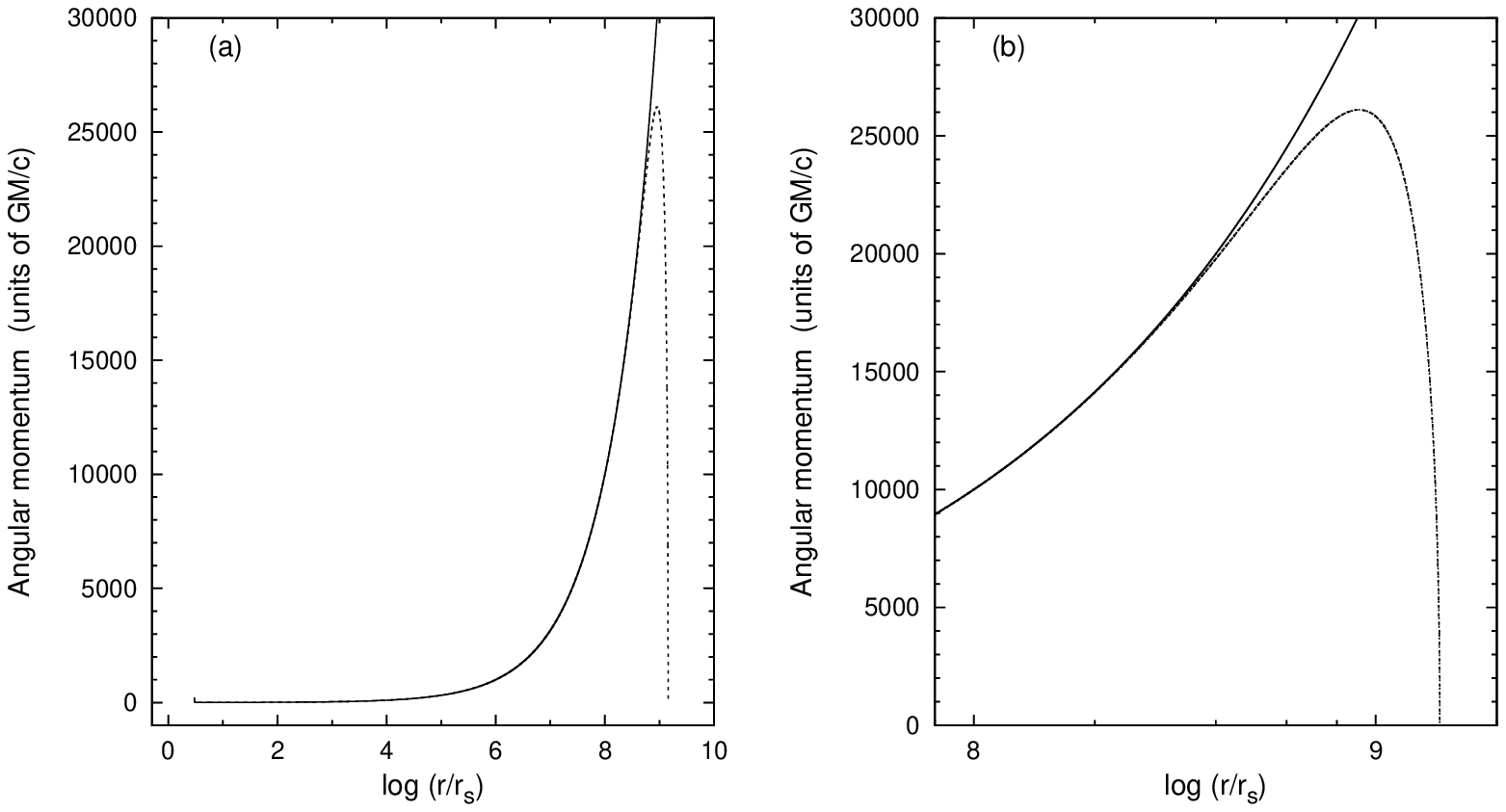}
\caption{Variation of angular momentum with $r$ for circular orbit trajectory. The short-dashed curve in Fig. 4a corresponds to $V_{\rm DS}$ with $\Lambda r^2_s = 1 \times 10^{-27}$ that coincides with the corresponding GR results. The solid curves in figures 4a and 4b are due to $V_{\rm SW}$. The long-dashed line in Fig. 4b corresponds 
to $V_{\rm DS}$ which coincides with the angular momentum profile corresponding to the PNP given in Eq. (22).
 }
\label{Fig4}
\end{figure}

\begin{figure}
\centering
\includegraphics[width=1.0\columnwidth]{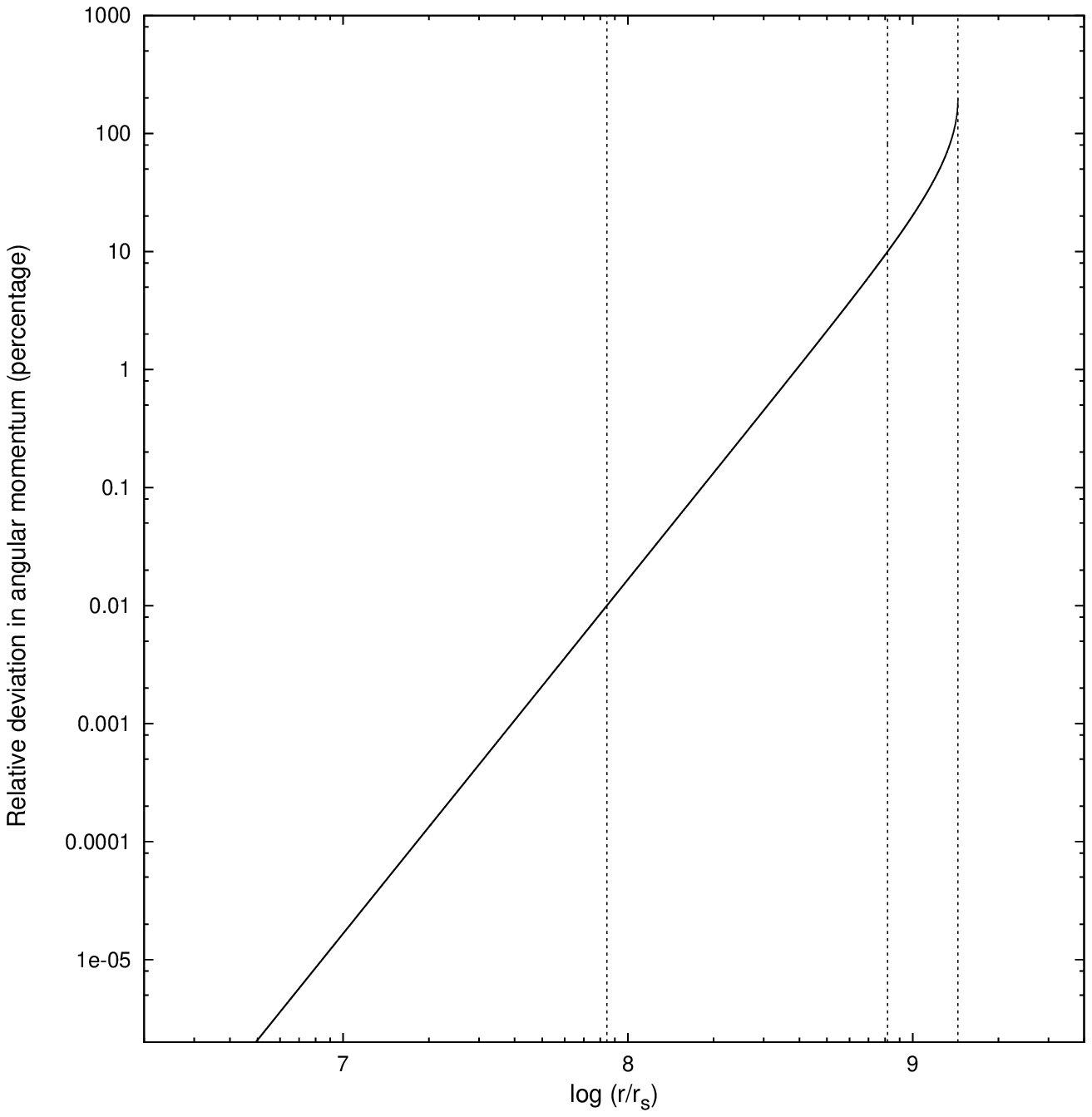}
\caption{Radial variation of the relative deviation (in percentage) in angular momentum. The two extreme vertical dashed lines, represented by $x_{\rm min}$ and $x_{\rm max}$, correspond to value of $\xi_{\imath}$ $\sim 0.01 \%$ and $\sim 200 \%$, respectively. The curve is truncated at $x_{\rm max}$ which also corresponds to the static radius. The vertical dashed line in the middle represents $r$ at which $\xi_{\imath}$ $\sim 10 \%$. 
 }
\label{Fig5}
\end{figure}

Figure 6 show the variation of conserved Hamiltonian $E^C_{\rm DS}$ with $r$, which is compared with the Hamiltonian corresponding to the PNP in Eq. (22) and the Hamiltonian corresponding to $V_{\rm SW}$. It needs to be noted that $E^C_{\rm DS}$ corresponding to $V_{\rm DS}$ exactly resembles the corresponding expression in general relativity. In the outer region of the test particle motion in circular orbit, where the effect of repulsive $\Lambda$ is prominent (figures 6b,c), there is a marked difference between the profile of conserved Hamiltonian corresponding to $V_{\rm DS}$ and that corresponding to the PNP in Eq. (22), especially at $r \, \gsim \, 5 \times 10^7 \, r_s$. This radius $r \sim 5 \times 10^7 \, r_s$ approximately resembles the location $x_{\rm min}$, the lower bound in $r$, at which the relative deviation 
$\left(\xi_{\imath}\right)$ in the Hamiltonian attains the value of $\sim 0.01 \%$ (see Fig. 6d). In fact, the value of $x_{\rm min}$ in this case is 
$\sim 5.3 \times 10^7 $. Around static radius $\left( \sim 10^9 \, r_s \right)$, the Hamiltonian corresponding to the PNP in Eq. (22) attains zero value (Fig. 6b), unlike the case of $E^C_{\rm DS}$. In the vicinity of cosmological horizon $\left(\sim 5.5 \times 10^{13} \, r_s \right)$, the Hamiltonian corresponding to the PNP in Eq. (22) diverges, showing a contrasting behavior with respect to the profile of $E^C_{\rm DS}$. $E^C_{\rm DS}$ after attaining a certain negative value becomes zero at a large outer radius, which is the outermost bound orbit (Fig. 6c). Due to the substantial effect of $\Lambda$, $E^C_{\rm DS}$ deviates significantly from that corresponding to $V_{\rm SW}$ at $r \, \gsim \, 5.5 \times 10^8 \, r_s$ (Fig. 6b), which precisely resembles the radius at which $\xi_{\imath}$ attains the value of $\sim 10 \%$ (Fig. 6d). Resembling Fig. 2b, here too, $x_{\rm max}$ is located at that radius where $\xi_{\imath}$ attains a value of $\sim 200 \%$. The value of $x_{\rm max}$ is then $\sim 7.5 \times 10^9$.

In the inner radii of the test particle in the circular orbit, where the effect of $\Lambda$ is negligible, $E^C_{\rm DS}$ shows appreciable deviation from the Hamiltonian profile corresponding to the PNP in Eq. (22). We do not show the corresponding profiles separately in this region, as already been furnished in [36].

\begin{figure}
\centering
\includegraphics[width=1.0\columnwidth]{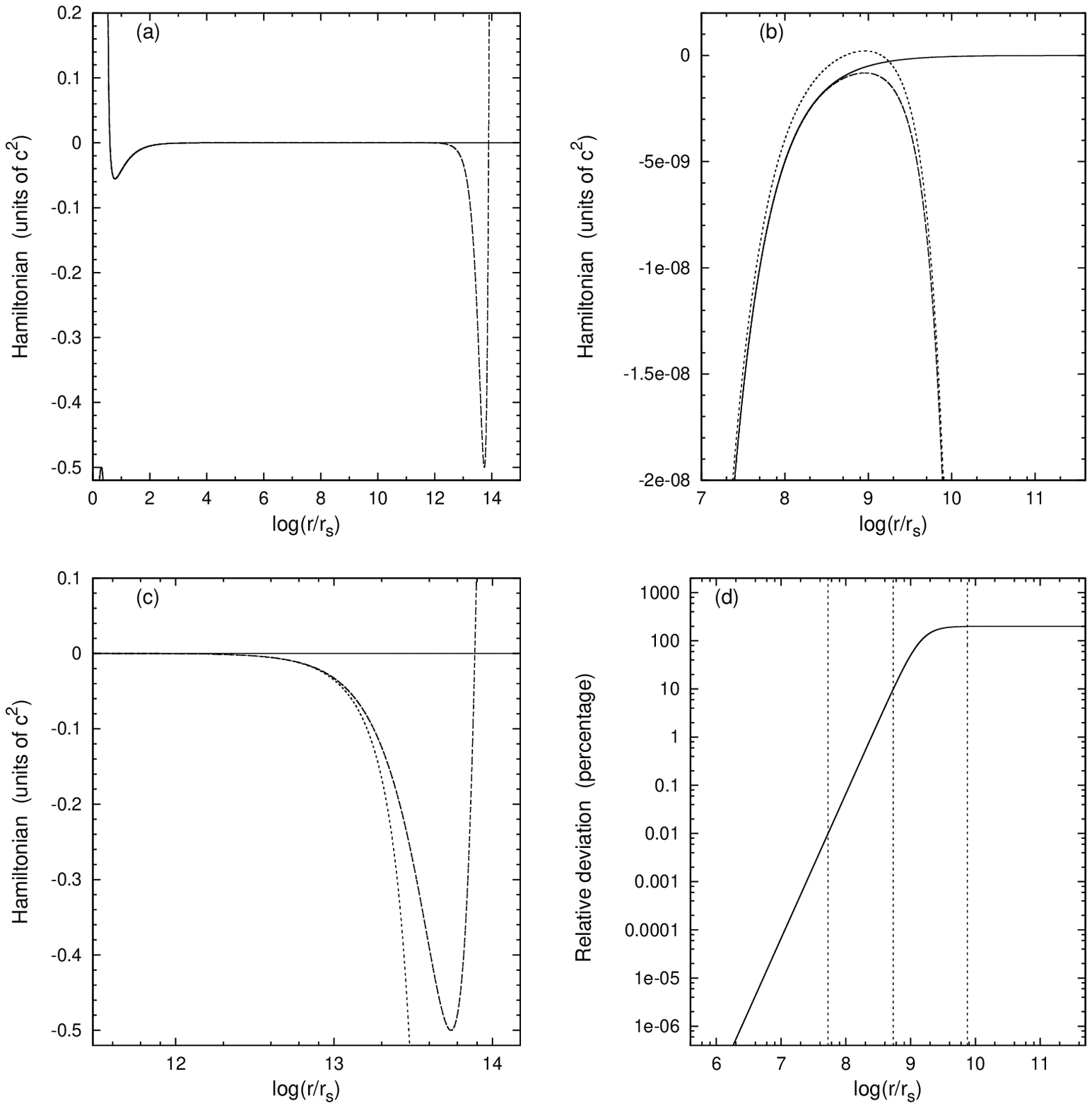}
\caption{Variation of Hamiltonian with $r$ for test particle in circular orbit corresponding to $\Lambda r^2_s = 1 \times 10^{-27}$. The solid and long-dashed lines in Fig. 6a are for the Hamiltonian corresponding to $V_{\rm SW}$ and $V_{\rm DS}$, respectively. The profile of $E^C_{\rm DS}$ corresponding to $V_{\rm DS}$ coincides with that in general relativity. The solid, long-dashed and short-dashed lines in figures 6b,c are for the Hamiltonian corresponding to $V_{\rm SW}$, $V_{\rm DS}$ and the PNP in Eq. (22), respectively. Figures 6b,c focus on the outer radii; near to the static radius and the cosmological horizon, respectively. Fig. 6d resembles Fig. 2b and Fig. 5, however, showing the the relative deviation in Hamiltonian (in percentage), as a function of radial distance $r$. 
 }
\label{Fig6}
\end{figure}

In Fig. 7a, instead of the profile of orbital angular velocity $\dot \Omega^{C}_{\rm DS}$, we depict the profile of corresponding angular 
frequency $\omega^{C}_{\rm DS} \left(= \dot \Omega^{C}_{\rm DS}/{2\pi} \right)$ in $r$, which truncates at the static radius at $\sim 1.4 \times 10^9 \, r_s$. At outer radii where the effect of $\Lambda$ is quite substantial, the value of angular frequency of the particle in circular orbit is negligible. Owing to which, $\omega^{C}_{\rm DS}$, its corresponding GR counterpart, angular frequency corresponding to $V_{\rm SW}$ and the angular frequency corresponding to the PNP in Eq. (22), all coincide. Hence, we do not show the angular frequency profile of the test particle motion in the circular orbit separately, at the outer radii, as well the corresponding relative deviation profile. However, in the inner radii, where the effect of $\Lambda$ is negligible, $\omega^{C}_{\rm DS}$ deviates from the corresponding GR result by a maximum of $6 \%$. Moreover, the angular frequency profile of the particle in the circular orbit corresponding to the PNP in Eq. (22) shows a huge deviation from the profile of $\omega^{C}_{\rm DS}$, as well as from the corresponding GR result, with an error margin of $\sim 75 \%$. The corresponding orbital angular velocity profiles in the inner radii have already been shown in [36].

\begin{figure}
\centering
\includegraphics[width=1.0\columnwidth]{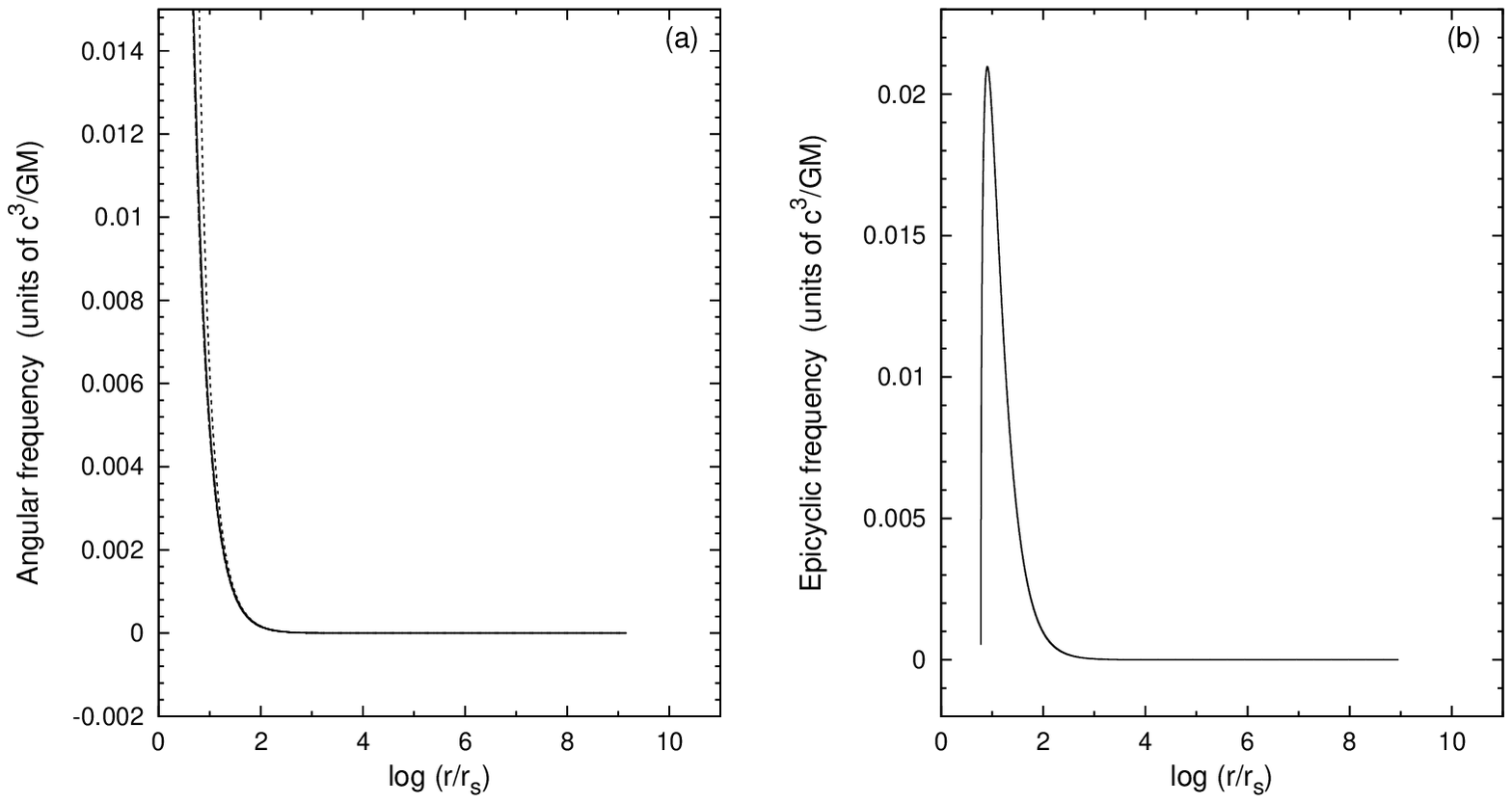}
\caption{The variation of the angular frequency and the epicyclic frequency with $r$ for the test particle in circular orbit. Solid, long-dashed and short-dashed curves in Fig. 7a depict angular frequency profiles corresponding to $V_{\rm DS}$ with $\Lambda r^2_s= 1 \times 10^{-27}$, the GR counterpart, and the PNP in Eq. (22) respectively. Note that $\omega^{C}_{\rm DS}$ coincides with the angular frequency profile corresponding to $V_{\rm SW}$, both in the inner as well as at the outer radii. Figure 7b depicts the variation of epicyclic frequency with $r$, corresponding to $V_{\rm DS}$
 } 
\label{Fig7}
\end{figure}

\subsection{Orbital perturbation and apsidal precession} 

As discussed in the previous subsection about the importance of the repulsive cosmological constant on the dynamical behavior of the particle orbits and their stability at the outer radii near the static radius, it is necessary for us to investigate the perturbative effects on the orbital dynamics, which would indeed be affected by the positive cosmological constant around the cosmological horizon. The perturbative effects would plausibly have substantial influence on the accretion flow stability in the outer regions in SDS background, in the expanding Universe. We follow our analysis by computing the epicyclic frequency for small perturbation of the particle orbit in circular trajectory. It should be reminded that till date there is no exclusive 
analytical relation of epicyclic frequency in full general relativity corresponding to SDS spacetime, unlike that in the case of Schwarzschild and Kerr BHs [38]. Henceforth, we evaluate the relation for epicyclic frequency using our correct SDS analogous potential $V_{\rm DS}$. We restrict ourselves in the equatorial plane of test particle orbit in a circular trajectory. We then perturb $r$ and $\phi$ and their derivatives accordingly: 
$r \to r + \delta r, \, \dot r \to \delta {\dot r},  \, \ddot{r} \to \delta \ddot{r}$ and
$\phi \to \phi + \delta \phi, \, {\dot \phi} \to {\dot \phi} \, {\vert}_C + \delta {\dot \phi}, \, \ddot{\phi} \to \delta {\ddot{\phi}} $ respectively. 
Using Eqs. (18), (19) and (20), the linearized perturbed equations of motion 
are then obtained as 

\begin{eqnarray}
\delta {\ddot{r}} = \delta r \, \left[{\dot \phi}^2 \, {\vert}_C + \left(r-2r_s - {\Lambda r^3}{3} \right) \left[\frac{2GM}{r^5} (r-4r_s) +
\frac{\Lambda c^2}{3r} \left(1+\frac{4r_s}{r} - \frac{5 \Lambda r^2}{3} \right) \right] \right] \, 
+ 2 {\dot \phi} \, {\vert}_C \, (r - 3r_s) \delta {\dot{\phi}}\, ,
\label{33}
\end{eqnarray}

\begin{eqnarray}
\delta {\ddot{\phi}} = -\frac{2 {\dot \phi} \, {\vert}_C }{r} \, \left(\frac{r - 3r_s}{r - 2 r_s - \frac{\Lambda r^3}{3} }  \right) \delta{\dot r}
\label{34}
\end{eqnarray}

and 

\begin{eqnarray}
\delta {\ddot{\theta}} = - {\dot \phi}^2 \, {\vert}_C \, {\delta \theta} \, ,
\label{35}
\end{eqnarray}

respectively. Here, ${\dot \phi} \, {\vert}_C$ is the orbital angular velocity of the particle motion in the circular orbit in equatorial plane. We concentrate ourselves in computing only the radial epicyclic frequency as it is more significant. Using Eqs. (33) and (34), with the following relations of perturbed quantities $\delta r = \delta r_0 \exp^{\imath \kappa t}$ and $\delta \phi = \delta \phi_0 \exp^{\imath \kappa t}$ for harmonic oscillations, where $\kappa$ is the radial epicyclic frequency and $\delta r_0$ and $\delta \phi_0$ are amplitudes (see [38]), we eventually obtain the expression for radial epicyclic frequency $\kappa$ corresponding to $V_{\rm DS}$ after rigorous algebra, given by  

\begin{eqnarray}
\kappa = \left(\frac{r-2r_s -\frac{\Lambda r^3}{3} }{r - 3r_s} \right)^{1/2} \, \sqrt{ \left[ \frac{GM}{r^5} (r-6r_s)(r-2r_s) 
- \frac{\Lambda c^2}{3r^2} \left[2 (r-4r_s)(2r-3r_s) - \frac{\Lambda r^3}{3} (4r-15r_s) \right] \right] } \, .
\label{36}
\end{eqnarray}

The relation in Eq. (36) reduces to that in Schwarzschild geometry with 
$\Lambda = 0$ whose value lies within a maximum error of just $\sim 6 \%$, as compared to the GR result. Although the radial epicyclic frequency $\kappa$ has been derived with SDS analogous potential, we anticipate that this relation could also be accurately used in full GR framework, as the effect 
of $\Lambda$ can only be realized at large outer radii from the central object, where the relativistic effect will itself be diminishing. As no other effective expressions for epicyclic frequency corresponding to SDS geometry is known, in Fig. 7b, we only show the variation of epicyclic frequency corresponding to  $V_{\rm DS}$ with $r$. Resembling $\omega^{C}_{\rm DS}$, the value of $\kappa$ at the outer radii is negligible. Consequently, the profile of epicyclic frequency corresponding to $V_{\rm DS}$ coincides with that corresponding to $V_{\rm SW}$, both in the inner as well as at the outer radii, and hence the corresponding relative deviation would also be negligible. In the inner radii, where the effect of 
$\Lambda$ is negligible, the PNP in Eq. (22) reduces to Paczy\'nski-Witta potential. The epicyclic frequency profile corresponding to this potential shows a huge deviation from the profile corresponding to $V_{\rm DS}$ by over an error margin of $\sim 75 \%$ (see [36]). 

Before analyzing the dynamics of orbital trajectory, we display the profile of the radial velocity $v_r$ (see Fig. 8) of the test particle falling radially towards the central BH in the SDS background. The expression of $v_r$, here, solely depends on the energy $E_{\rm DS}$ of the test particle motion. In figures 8a,b we show the profiles of $v_r$ corresponding to $V_{\rm DS}$, $V_{\rm SW}$, and the PNP in Eq. (22), for the test particle with $E_{\rm DS} \sim 0$ and $E_{\rm DS} \sim 0.5$, respectively. Expression for $v_r$ corresponding to $V_{\rm DS}$ coincides with the GR expression. 
In figures 8c,d, we depict the detailed nature of the corresponding profiles at the outer radii which is our region of interest, where the effect of $\Lambda$ is significant. The figures show that $v_r$ profiles corresponding to the PNP in Eq. (22) distinctly differ; rather show opposite behavior from $v_r$ corresponding to $V_{\rm DS}$ at the outer radii, at $r \, \gsim \, 6 \times 10^{12} \, r_s$. For semi-relativistic/relativistic test particle energy, the radial velocity corresponding to the PNP in Eq. (22) even surpasses velocity of light at that radii, thus showing physically inconsistent behavior (figures 8b,d). In figures 8e,f, we 
show the profiles of relative deviation $\left(\xi_{\imath} \right)$ in $v_r$, corresponding to $E_{\rm DS} \sim 0$ and $E_{\rm DS} \sim 0.5$, respectively. $x_{\rm min}$ and $x_{\rm max}$ corresponding to $v_r$ for $E_{\rm DS} \sim 0$ are located at $r \sim  1 \times 10^{8} \, r_s$ and $r \sim 5.8 \times 10^{10} \, r_s$, at which $\xi_{\imath}$ attains the value of $\sim 0.01 \%$ and $\sim 200 \%$, respectively. However, the relative deviation corresponding to semi-relativistic/relativistic test particle energy (Fig. 8f) show marked difference as compared to test particle motion in the low energy limit, unlike that in Fig. 2b. For the semi-relativistic/relativistic energy, $v_r$ corresponding to $V_{\rm DS}$ deviates from that corresponding to $V_{\rm SW}$ at much outer radii. $x_{\rm min}$ and $x_{\rm max}$ thus move further outward attaining the value of $\sim 7 \times 10^{11}$ and $\sim 5.5 \times 10^{13}$, where $\xi_{\imath}$ is $\sim 0.01 \%$ and $\sim 200 \%$, respectively. In Fig. 8f, the curve gets truncated at $x_{\rm max}$ which is also the cosmological horizon.  

The dynamics of orbital trajectory can be obtained from Eq. (17), which is identical to that obtained in general relativity, implying that the nature of trajectory of particle orbit obtained with $V_{\rm DS}$ would be similar to that in general relativity. In Fig. 9 we show the profile of ${dr}/{d\Omega}$ with $r$ corresponding to $\Lambda r^2_s = 1 \times 10^{-27}$. The profiles indicate that for semi-relativistic/relativistic test particle energy, ${dr}/{d\Omega}$ corresponding to $V_{\rm DS}$ deviate from ${dr}/{d\Omega}$ corresponding to $V_{\rm SW}$ at much outer radii (figures 9b,d), as compared to test particle energy in the low energy limit 
(figures 9a,c). ${dr}/{d\Omega}$ profiles corresponding to PNP in Eq. (22) also show different behavior, from that corresponding to $V_{\rm DS}$.  

\begin{figure}
\centering
\includegraphics[width=1.0\columnwidth]{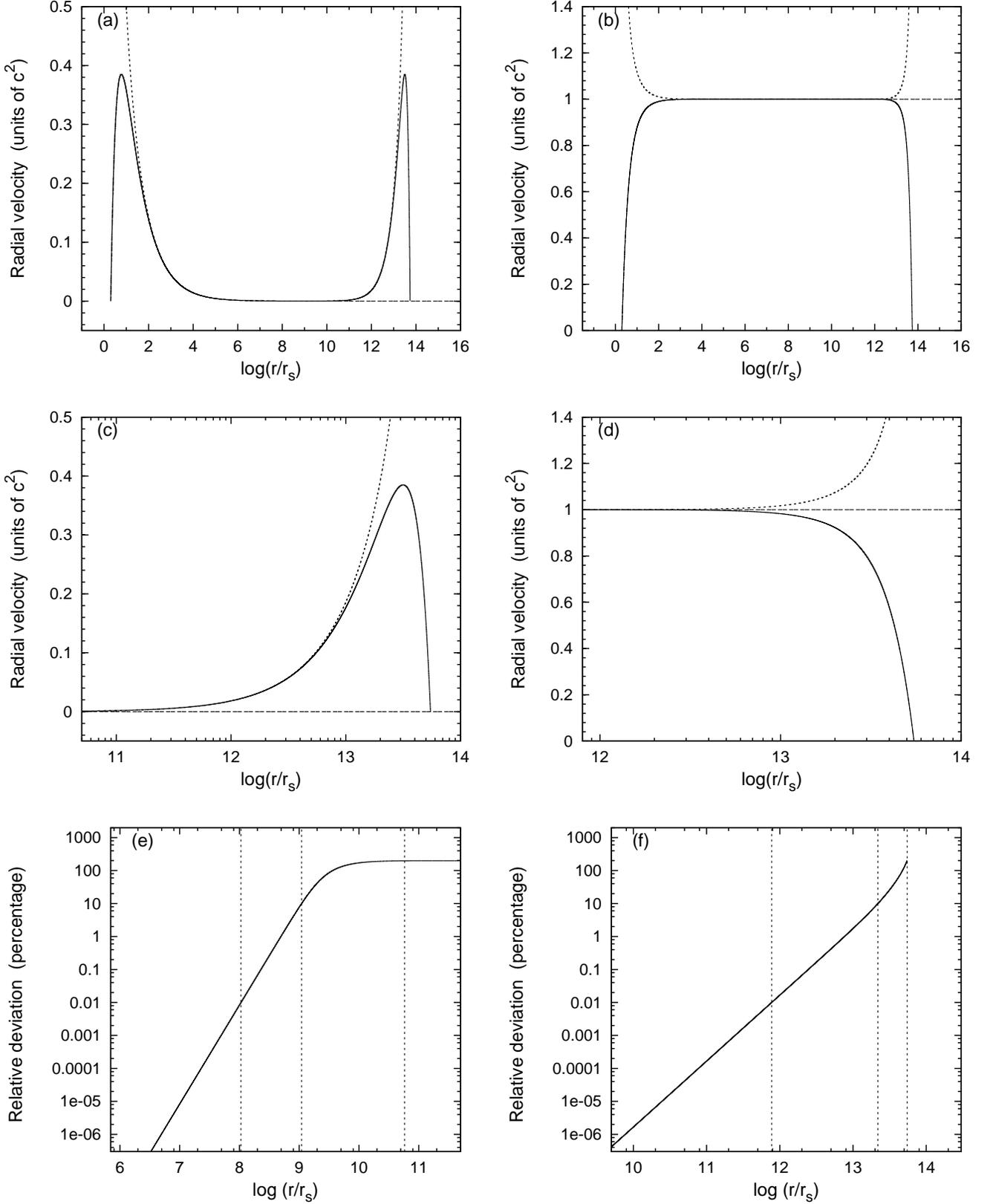}
\caption{Variation of radial velocity of the test particle falling radially towards the central BH. Solid, long-dashed and short-dashed curves in Fig. 8a are for $v_r$ corresponding to $V_{\rm DS}$ with $\Lambda r^2_s = 1 \times 10^{-27}$, $V_{\rm SW}$ and PNP in Eq. (22), respectively, for test particle with $E_{\rm DS} \sim 0$. Figure 8b is similar to that in 8a, however, for $E_{\rm DS} \sim 0.5$. In figures 8c,d which resemble figures 8a,b, the detailed nature of the corresponding profiles at the outer radii is depicted. In figures 8e,f we show the relative deviation (in percentage) in $v_r$, corresponding to $E_{\rm DS} \sim 0$ and $E_{\rm DS} \sim 0.5$, respectively. The vertical lines in the figures are same to those in Fig. 2b. $E_{\rm DS}$ is expressed in units of $c^2$. 
 }
\label{Fig8}
\end{figure}

\begin{figure}
\centering
\includegraphics[width=1.0\columnwidth]{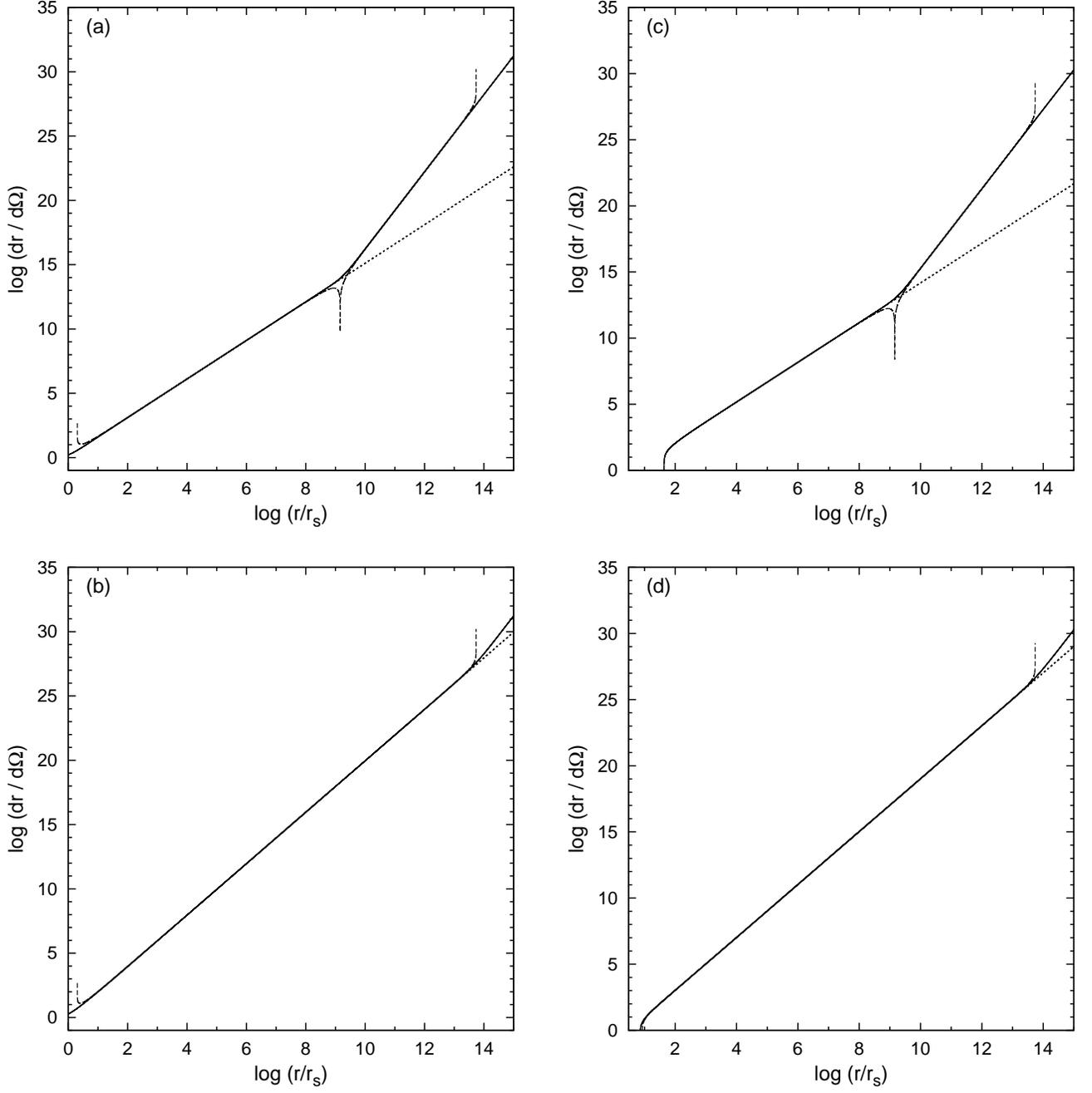}
\caption{Variation of ${dr}/{d\Omega}$ in radial direction $r$ corresponding to $\Lambda r^2_s = 1 \times 10^{-27}$. The solid, long-dashed and short-dashed curves in all the figures correspond to $V_{\rm DS}$, PNP in Eq. (22), and $V_{\rm SW}$. Figures 9a and 9c are for $E_{\rm DS} \sim 0$ corresponding to angular momentum $\lambda_{\rm DS} =(1.1, 9.5)$, respectively.  Figures 9b and 9d resemble figures 9a,c, however for $E_{\rm DS} \sim 0.5$. $E_{\rm DS}$ and $\lambda_{\rm DS}$ are expressed in units of $c^2$ and $GM/c$, respectively. 
 }
\label{Fig9}
\end{figure}

\begin{figure}
\centering
\includegraphics[width=1.0\columnwidth]{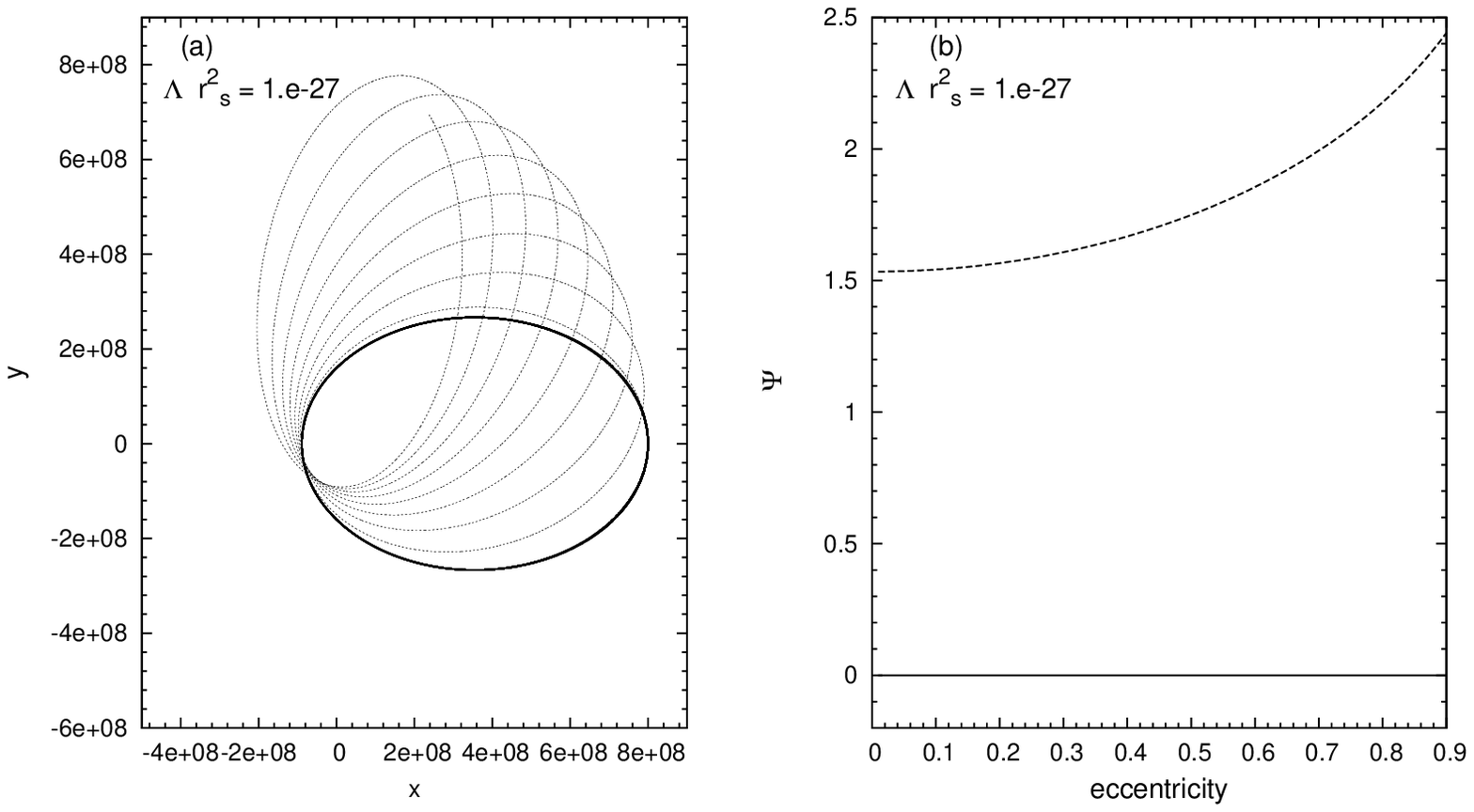}
\caption{The profile of elliptical like trajectory of particle orbit in equatorial plane due to $V_{\rm DS}$ with $\Lambda r^2_s = 1 \times 10^{-27}$. In Fig. 10a, the solid line is due to $V_{\rm SW}$ and the dotted line for $V_{\rm DS}$. The particle trajectory coincides with that in general relativity. Figure 10b shows the profile of perihelion advancement $\Psi$ as a function of eccentricity corresponding to $V_{\rm DS}$ for $r_a = 8 \times 10^8 \, r_s$. The solid and long-dashed curves correspond to  $V_{\rm SW}$ and $V_{\rm DS}$ respectively. 
 }
\label{Fig10}
\end{figure}

In Fig. 10a, the elliptical-like trajectory of particle orbit in the $x-y$ plane due to $V_{\rm DS}$, along with that due to $V_{\rm SW}$, obtained from the equations of motion are shown. The particle starts with a tangential initial velocity of $1.58114 \times 10^{-5} \, c$ from an apoapsis $r_a = 8 \times 10^8 \, r_s$ attaining a constant eccentricity of magnitude $0.8$. We obtained the plot of elliptical trajectory using Cartesian transformation adopting the method of Euler-Cromer algorithm which preserves energy conservation. The distinct precessional effect can be seen around $\sim 80 \, {\rm kiloparsec}$. This is precisely the radius at which the $\xi_{\imath}$ corresponding to the potential (Fig. 2b) becomes greater than $\sim 10 \%$. In Fig. 10b, we show the apsidal precession or the perihelion advancement $\Psi$ [given by Eq. (37)] as a function of eccentricity which is identical to the GR results. This also guarantees that the photon orbit trajectory obtained with 
$V_{\rm DS}$ would resemble that in general relativity, which is independent of the cosmological constant. 

\begin{eqnarray}
\Psi = \Pi - \pi \equiv \int_{r_p}^{r_a} \frac{d \phi}{dr} \, dr - \pi  \, ,
\label{37}
\end{eqnarray}

where $\Pi$  is the usual half orbital period of the test particle. $r_p$ and $r_a$ are periapsis and apoapsis of the orbit respectively. 

\subsection{Stability and boundedness of circular orbit} 

We can obtain the last stable circular orbit using potential $V_{\rm DS}$ with the condition ${d\lambda^{C}_{\rm DS}}/{dr} = 0$ or equivalently 

\begin{eqnarray}
3 r_s r - 18 r^2_s - 4 \Lambda r^4 + 15 \Lambda r^3 r_s = 0 \, , 
\label{38}
\end{eqnarray}

which is exactly similar to that obtained in full general relativity. Thus, $V_{\rm DS}$ would accurately reproduce the last stable circular orbit in general relativity. Also the last bound circular orbit is obtained from Eq. (28) with 
$E^C_{\rm DS} =0$, and by virtue, is similar to that in full general relativity. The equivalent relation is 

\begin{eqnarray}
9 r r_s - 36 r^2_s + 6 \Lambda r^4 - 12 \Lambda r^3 r_s - \Lambda^2 r^6 = 0 \, . 
\label{39}
\end{eqnarray}

For cosmological parameter ${\zeta} = 0$, the familiar circular orbit stability limit and the last bound circular orbit for the Schwarzschild metric is recovered. Unlike the case in the Schwarzschild geometry, circular orbits here become unstable at both inner and outer radii. Equation (38) gives two real positive roots. One of the root is the last stable circular orbit in the inner radii resembling the usual Schwarzschild case. The other root is the maximum possible circular stable orbit $r^{\rm SC}_{\rm max}$ at the outer radii which is located near to the static radius.
The location of $r^{\rm SC}_{\rm max}$ is $\sim 0.88 \times 10^9 \, r_s$, whereas the corresponding static radius is located at $\sim 1.4 \times 10^9 \, r_s$. Resembling the case for stable circular orbit, Eq. (39) gives two real positive roots, corresponding to the marginally bound circular orbit in the inner radii and the maximum possible bound circular orbit $r^{\rm BC}_{\rm max}$ at the outer radii. Inner marginally bound circular orbit is the usual limit corresponding to Schwarzschild geometry. PNP in Eq. (22), too, accurately reproduces both marginally stable and bound orbits corresponding to SDS geometry. However, it needs to be noted that the test particle in SADS background with negative $\Lambda$ does not yield any outermost stable or outermost bound circular orbit. 

Although we have studied the entire test particle dynamics with $\Lambda r^2_s = 1\times 10^{-27}$ for BH of $\sim 10^9 M_{\odot}$, however, in Fig. 11, we will depict the variation of $r^{\rm SC}_{\rm max}$ and $r^{\rm BC}_{\rm max}$ with $\Lambda r^2_s$, and the corresponding values of the dynamical variables for the test particle motion in circular orbit, in order to see the effect of $\Lambda$ on the stability of the orbits. With the decrease in value of $\Lambda r^2_s$, both $r^{\rm SC}_{\rm max}$ and $r^{\rm BC}_{\rm max}$ move further outwards. For all values of $\Lambda r^2_s$, $r^{\rm BC}_{\rm max}$ is located at much outer radii as compared to that of $r^{\rm SC}_{\rm max}$. We show the variation of $\lambda^{C}_{\rm DS}, E^C_{\rm DS}$ and $\omega^{C}_{\rm DS}$ along $r^{\rm SC}_{\rm max}$ corresponding to various values of $\Lambda r^2_s$ in figures 11b,c,d, respectively, which are identical to the GR results, as well as corresponding to the PNP in Eq. (22). On the other hand, the corresponding dynamical variable do not furnish any real values along $r^{\rm BC}_{\rm max}$. On the contrary, the values of the dynamical variables corresponding to $V_{\rm DS}$ in the innermost stable and innermost bound orbits where the effect of $\Lambda$ is negligible, differ considerably from that corresponding to the PNP in Eq. (22), as discussed in [36].   
   
\begin{figure}
\centering
\includegraphics[width=1.0\columnwidth]{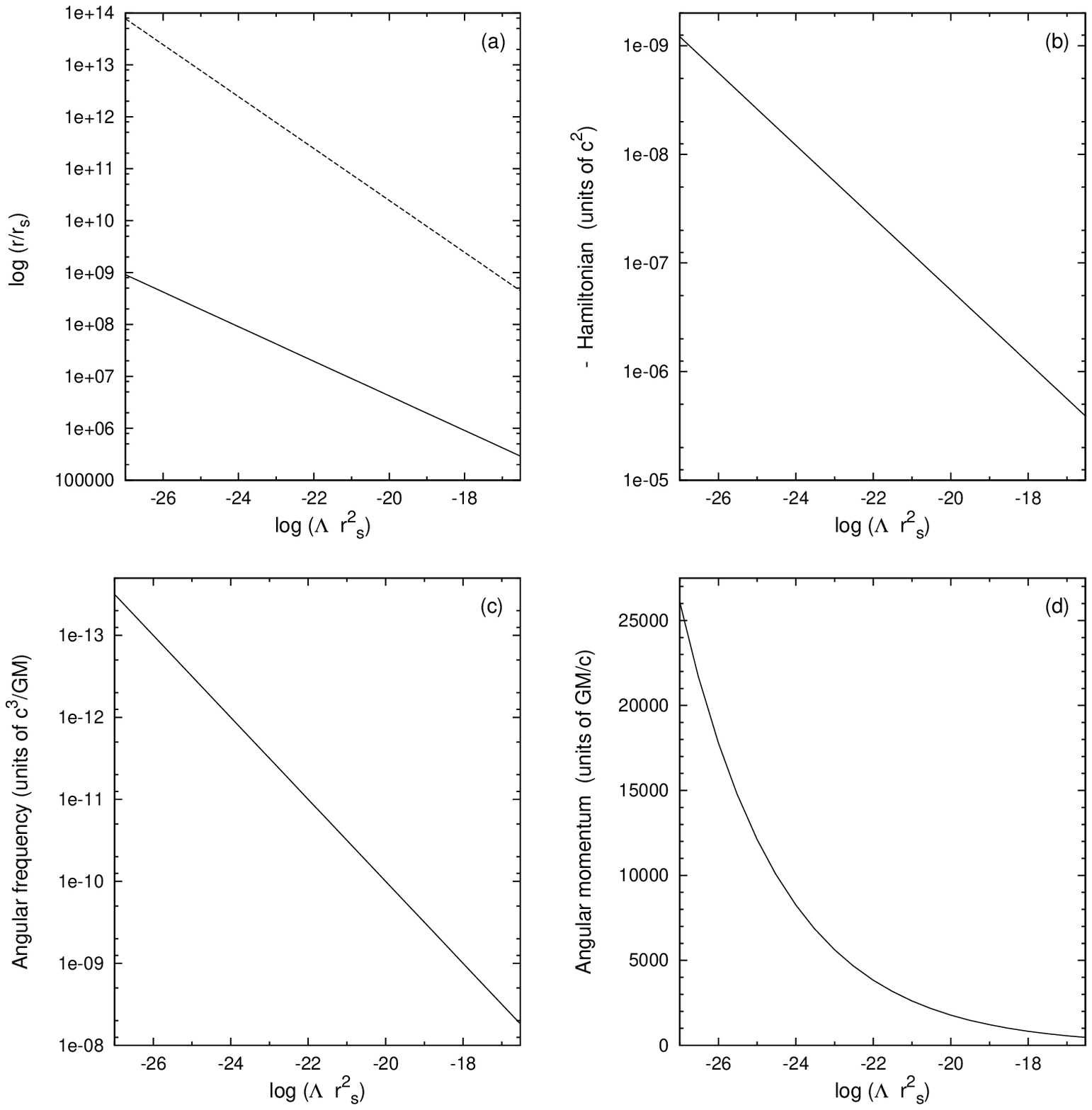}
\caption{Variation of the location of outermost stable circular and outermost bound circular orbit with repulsive $\Lambda$, obtained with $ V_{\rm DS}$. The solid and dashed lines in Fig. 11a show the variation of $r^{\rm SC}_{\rm max}$ and $r^{\rm BC}_{\rm max}$ with $\Lambda$, respectively. Figures 11b,c,d show the variation of Hamiltonian, angular momentum and angular frequency along $r^{\rm SC}_{\rm max}$ with corresponding values of $\Lambda$, obtained with $V_{\rm DS}$. The y-axis in Fig. 11a is in logarithmic-scale.
 }
\label{Fig11}
\end{figure}

%In Table 1, we furnish the exact estimate of the values of the dynamical variables for $\Lambda r^2_s=1 \times 10^{-27} $ at the outermost stable and bound orbit, corresponding to our potential which are identical with that in general relativity and PNP of Eq. (23).

%\begin{table*}[htbp]
%\large
%\centerline{\large Table 1}
%\centerline{\large Values of dynamical variables at $r^{\rm SC}_{\rm max}$ and $r^{\rm BC}_{\rm max}$}
%\begin{center}
%\begin{tabular}{ccccccccc}
%\hline
%\hline
%\noalign{\vskip 2mm}
%$\Lambda r^2_s$  & $r^{SC}_{\rm max}$ & $r^{BC}_{\rm max}$ & $\lambda^{C}_{\rm SC}$  & $E^{C}_{\rm SC}$  & ${\dot \Omega}^{C}_{\rm SC}$ & $\kappa_{\rm SC}$  \\
%\hline
%\hline
%\noalign{\vskip 2mm}
%$1\times 10^{-27}$  &  9.086e8  & 7.746e13  & 2.610e4 & -8.255e-10  &  3.162e-12  &  ${\rm im}$ \\
%\hline
%\end{tabular}
%\end{center}
%\end{table*}

%As mentioned earlier, we do not obtain any real values of the corresponding dynamical variables at the outermost bound orbit. From Table 1, it is seen that the epicyclic frequencies along $r^{\rm SC}_{\rm max}$ with corresponding values of $\Lambda$ do not follow either a monotonically increasing or decreasing pattern, unlike other dynamical variables. Moreover, we do not obtain any real value of epicyclic frequency for $\Lambda r^2_s = 1\times 10^{-27}$. 

For supermassive BHs (SMBHs) in AGNs or quasars with $\left(10^6 M_{\odot} \, \lsim \,  M_{\rm BH} \, \lsim \, 10^9 M_{\odot} \right)$, the current accepted value of $\Lambda$ yields a maximum possible radius of the outermost stable circular orbit in the range of [$\sim (0.9 - 88)$] kiloparsec. This also implies that the outer edge of the largest Keplerian accretion disk in AGNs/Quasars will never exceed length-scale of kiloparsecs. The instability of circular orbits at large outer radii can have serious astrophysical consequences. In the next section, we analyze the influence of $\Lambda$ on a simplistic astrophysical phenomenon, Bondi accretion rate. 

\subsection{A comparative analysis}

In the last section, we have exemplified that the test particle dynamics corresponding to $V_{\rm DS}$ are mostly identical to the corresponding GR results. This is due to the fact that geodesic equations obtained from the Lagrangian corresponding to $V_{\rm DS}$ are identical to the GR results, in the low energy limit. The derived potential exactly reproduces the GR results like horizon properties, orbits including apsidal precession, angular momentum, energy, radial velocity of the test particle and also the temporal properties.  

In the inner radii, where the effect of $\Lambda$ is negligible, the profile of $V_{\rm DS}$ and particle dynamics corresponding to it essentially coincide with 
those of $V_{\rm SW}$. However, as seen from the analysis in the earlier subsections, the effect of $\Lambda$ starts to become prominent in the outer region from an approximate radius $x_{\rm min}$, at which $\xi_{\imath}$ corresponding to most of the quantities attain a value of $\sim 0.01 \%$. With $\Lambda r^2_s= 1 \times 10^{-27}$, for most of the quantities, the corresponding values of $x_{\rm min}$ is found out to be approximately in the 
range of $\sim  \left(5 \times 10^7 - 1 \times 10^8 \right) \, r_s$. This is approximately the region from where $V_{\rm DS}$ and most of its corresponding dynamical quantities starts to deviate effectively, from that corresponding to $V_{\rm SW}$. For $M_{\rm BH} \sim 10^9 \, M_{\odot}$, this gives a radius of $\sim (5-10)$ kiloparsec. This implies that in massive galaxies, beyond a scale of few  kiloparsecs, the influence of repulsive cosmological constant may not be neglected. Nevertheless, the most significant effect of $\Lambda$ occurs from around a radius of $\gsim \, \left(5 \times 10^8 - 1 \times 10^9 \right) \, r_s$, where $\xi_{\imath}$ corresponding to most of the dynamical quantities attain a value of  $\gsim \, 10 \%$. With BH of $\sim 10^9 \, M_{\odot}$, this gives a radius of $\sim (50-100)$ kiloparsec. 
This is approximately the region where $V_{\rm DS}$ and its corresponding dynamical quantities 
largely deviate from that corresponding to Schwarzschild case. This is also the radius where strong orbital precessional effect in the SDS background 
can be seen (Fig. 10).   

On the other hand, the results obtained with $V_{\rm DS}$ differ significantly from the corresponding results of the PNP in Eq. (22), both in the inner radii as well as at the outer radii. In the inner radii $(\lsim 100 \, r_s)$, where the effect of $\Lambda$ is negligible, the PNP in Eq. (22) reduces to that of 
Paczy\'nski-Witta potential, and the particle dynamics corresponding to this PNP deviates largely from that obtained with $V_{\rm DS}$, or in general relativity. The details of which have been elucidated in [36]. However, in the outer region where the effect of $\Lambda$ is predominant, especially around the static radius and beyond $\left(\gsim \, 10^9 \, r_s \right)$, the  PNP in Eq. (22) as well as the profile of conserved Hamiltonian show prominent deviations from that of $V_{\rm DS}$ or that in general relativity, for test particle motion in circular orbit; sometimes showing  contrary behavior near to the cosmological horizon (Fig. 6). In fact, the Hamiltonian corresponding to the PNP in Eq. (22) starts deviating from that corresponding to $V_{\rm DS}$ or GR result, from an approximate location 
of $x_{\rm min}$. Moreover, the PNP in Eq. (22) gives opposite trend for the test particle radial velocity (figures. 8c,d), as compared to $V_{\rm DS}$ or in general relativity, at  $r \, \gsim \, 6 \times 10^{12} \, r_s$.
However, in general, for circular motion, the velocity dependent part of $V_{\rm DS}$ vanishes and it gives results close to that predicted by Eq. (22). Therefore, major differences are expected for non-circular motions and in such situations the PNP proposed in [34] does not yield consistent results with the GR (with $\Lambda$) and hence with potential $V_{\rm DS}$. 

Thus the PNP in Eq. (22) does not reproduce GR features with modest accuracy, unlike that with $V_{\rm DS}$. This occurs owing to the fact that $V_{\rm DS}$ has been derived based on the prerequisite, that, any analogous potential of corresponding GR geometry should reproduce identical or nearly-identical geodesic equations of motion. However, the only GR notion that the PNP in Eq. (22) tries to mimic is to reproduce last stable circular orbits and marginally bound circular orbits (both innermost and outermost). 

\section{Effect of $\Lambda$ on Bondi accretion rate}  

Bondi accretion is a spherically symmetric steady gaseous accretion onto a compact star [39]. Although Bondi accretion flow is more hypothetical, yet it has serious astrophysical relevance. It is generally accepted that low luminous AGNs or low excitation radio galaxies (LERGs) are presumably powered by radiatively inefficient hot mode accretion, accreting gaseous plasma from hot X-ray emitting phase of interstellar medium (ISM) or intergalactic medium (IGM) [40,41]. Consequently the accretion paradigm is quasi spherical in nature, accreting at near `Bondi accretion rate' from ISM/IGM. In the expanding Universe, Bondi accretion rate itself would be plausibly influenced by repulsive $\Lambda$, thus would explicitly govern the accretion dynamics [19].  

Dynamically Bondi solution is transonic in nature with two critical radii or sonic radii ($r_c$) at which the radial velocity $v_r$ exactly equals to sound speed $a$ in the limits $r \rightarrow {\infty}$ and $r \rightarrow 0$ respectively. The limit $r \rightarrow {\infty}$ in astrophysical sense represents the location of the ambient medium (ISM/IGM). As usual, the flow is considered to be locally adiabatic, with $P \propto \rho^\gamma$ and $a = \sqrt{\gamma P/\rho}$. $P$ is the gas pressure, $\rho$ is the density of the accreting gas and $\gamma$ the adiabatic index.
The mass accretion rate here is then defined by the relation 

\begin{eqnarray}
\vert \dot M \vert =  4 \pi r^2 \rho \, v_r \, .  
\label{40}
\end{eqnarray}

Without going into the detailed physics of Bondi flow which has been extensively discussed in literature (for e.g. [39,42]), we simply furnish 
the relations at critical point $r_c$, obtained from Eq. (40) and radial momentum conservation equation. Using the condition at critical radius, i.e., at 
$r = r_c$, $v_r \vert_c = a_c$ and using $V_{\rm DS}$, instead of usual Newtonian gravity, the relations at critical point $r_c$ are then given by 

\begin{eqnarray}
\frac{v^2_r \vert_c}{2} +  \frac{a^2_c}{\gamma -1} + V_{\rm DS} \vert_c = 
\frac{a^2_{\infty}}{\gamma-1}
\label{41}
\end{eqnarray}

and 

\begin{eqnarray}
\frac{2 a^2_c}{r_c} - \left. \frac{d V_{\rm DS/ADS}}{dr} \right\vert_c = 0 \, ,
\label{42}
\end{eqnarray}

respectively. Solving these two equations we obtain $r_c$ and corresponding $a_c$ in terms of $a_{\infty}$, where $a_{\infty}$ is the sound speed at 
$r \rightarrow {\infty}$. Using this, the density at $r_c$ is given by $\rho_c =  \rho_{\infty} \left(\frac{a^2_c}{a^2_{\infty}} \right)^{1/(\gamma-1)}$, where, $\rho_{\infty}$ is the density at $r \rightarrow {\infty}$. The transonic accretion rate is then given by 

\begin{eqnarray}
\vert \dot M_t \vert = 4 \pi r^2_c \, \rho_c \, a_c \, .  
\label{43}
\end{eqnarray}

$\vert \dot M_t \vert$ can then be expressed in terms of $a_{\infty}$ and $\rho_{\infty}$, which is defined as `Bondi accretion rate' $(\vert \dot M_B \vert )$ in SDS background with repulsive cosmological constant in presence of $V_{\rm DS}$. 
Assuming the accreting plasma to be of purely ionized hydrogen, we define $a_{\infty} =\sqrt{\gamma k_B T_{\infty}/m_p}$ and $\rho_{\infty} = n_p \vert_{\infty} \, m_p$, where $T_{\infty}$ and $n_p \vert_{\infty}$ are the temperature and the number density of the accreting gas at $r \rightarrow {\infty}$. $m_p$ and $k_B$ are the usual proton mass and Boltzmann constant, respectively. With these relations $\vert \dot M_B \vert$ in presence of $V_{\rm DS}$ can then be explicitly expressed in terms of $T_{\infty}, \rho_{\infty}, \Lambda, \gamma$ 

For Newtonian gravitational potential the usual relation of $\vert \dot M_B \vert = 4 \pi \, q(\gamma) \, G^2 M^2 \rho_{\infty} /{a^3_{\infty}} $, where $q(\gamma) = \frac{1}{4} \left(\frac{2}{5-3\gamma} \right)^{(5-3\gamma)/(2\gamma-2)}$. $T_{\infty}$ and $\rho_{\infty}$ in astrophysical sense are the temperature and density of the ambient medium, like hot X-ray emitting phase of ISM/IGM. In Fig. 12, we study the variation of $\vert \dot M_B \vert$ and sonic radius $r_c$ with $T_{\infty}$
and $\gamma$ for $\Lambda \sim 10^{-56} \rm cm^{-2}$. It is found that the 
current accepted value of cosmological constant does not have much impact on the sonic radius as well as the Bondi accretion rate for large values of ambient 
temperature. However, it is found  that there is a moderate increase in the value of $\vert \dot M_B \vert$ as compared to the usual value of $\vert \dot M_B \vert$ with $\Lambda = 0$, for an ambient temperature $T_{\infty} < 10^4 K$, and for smaller values in $\gamma$ ($\lsim \, 1.5$). The sonic radius of the flow, too, shift little inwards in SDS back ground as that compared to $\Lambda = 0$, corresponding to $T_{\infty} < 10^4 K$ and $\gamma \, \lsim \, 1.5$. 
This impact of $\Lambda$ on Bondi accretion rate 
might have interesting consequences on the advective accretion dynamics, especially in low luminous AGNs or in LERGs. 

\begin{figure}
\centering
\includegraphics[width=1.0\columnwidth]{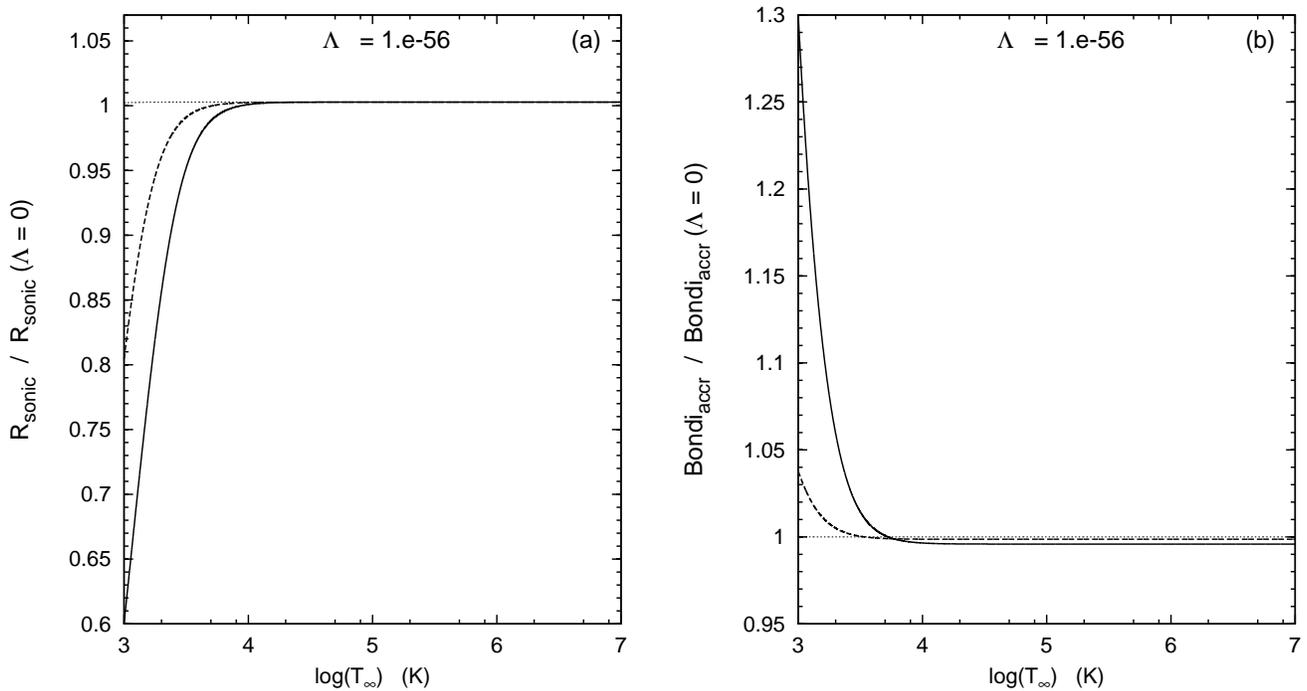}
\caption{Variation of sonic radius and  Bondi accretion rate with temperature $T_{\infty}$ obtained with $V_{\rm DS}$, for $\Lambda \sim 10^{-56} \rm cm^{-2}$. 
Figure 12a shows the variation of the ratio of sonic radius in the SDS background to the sonic radius with $\Lambda = 0$, in temperature. Figure 12b is similar to 12a, but for the ratio of Bondi accretion rate in SDS background to Bondi accretion rate with $\Lambda = 0$. Solid, long-dashed and dotted curves in the figures 12a,b are for adiabatic constant $\gamma = (4/3, 1.5, 1.66)$, respectively. BH mass is chosen to be $\sim 10^9 M_{\odot}$. 
 }
\label{Fig12}
\end{figure}

\section{Discussion}

The influence of positive cosmological constant is important in understanding the kinematics of the kiloparsecs-scale regions and beyond, in the local-galaxies. Any observational inference on the possible dependence of the local-scale structure of the galaxies on $\Lambda$ would be very intriguing, for which, Newtonian analogous potential would be of remarkable use. Here, we have obtained modified Newtonian analogous potential corresponding to SDS/SADS geometry which reproduces almost all of the GR features with remarkable accuracy. We have investigated the dynamical behavior of 
the test particle motion in SDS background for a typical BH mass $M_{\rm BH} \sim 10^9 M_{\odot} $ in accordance with the BH mass in many AGNs/quasars. For such massive black hole system, the non-negligible effects of cosmological constant $\Lambda$ start to occur at radius $\, x_{\rm min} \sim \, (5-10) \, {\rm kiloparsec}$ where the relative deviation between the quantities of SDS and Schwarzschild geometries attain a value of $\sim 0.01 \%$. The significant effects (at the level of $10 \%$ and above over those due to pure Schwarzschild geometry) of $\Lambda$ occur at $\, r\gsim \, (50-100) \, {\rm kiloparsec}$ and consequently the local astrophysical kinematics in many massive AGNs/quasars would be strongly influenced by $\Lambda$ beyond a distance of few kiloparsecs.

`Low excitation radio galaxies', which are often massive elliptical galaxies [43] are powered by the gaseous accretion directly from the X-ray emitting phase of ISM or either from the hot X-ray halos surrounding the galaxy or from the hot phase of the IGM [40,41], in which case the accretion flow region may well exceed 
parsecs-scale, extending even to kiloparsecs-scale. Quasar accretion disk may be susceptible to gravitational instability if the disk is self-gravitating and massive, which may trigger massive star formation in the outer regions extending beyond parsecs-scale [44]. Effects of $\Lambda$ in the outer regions of these kiloparsecs-scale accretion disk, especially on their dynamical instabilities, would then have interesting consequences in the expanding Universe. As remarked by [15] that positive cosmological constant could also influence the accretion process onto primordial BHs in the early epoch of the expanding Universe and can also induce strong collimation effects in astrophysical jets. Over kiloparsecs to megaparsecs-scale jets are, in fact, observed [45,46] in quasars and radio galaxies whose dynamics and structure would likely to be influenced by the effect of repulsive $\Lambda$ in the expanding Universe.  

Seemingly, those are not the only plausible ways in which a positive cosmological constant can, in principle, influence the astrophysical processes on the 
kiloparsecs-scale regions and beyond, in the local-galaxies in SDS background. Recently [47,48], it has been suggested that nuclear spirals can be one of the mechanisms to power central SMBHs by transferring gas from galactic-scales (over kiloparsec-scales) to the central parsec region. In such circumstances $\Lambda$ can influence the feeding process where the mass inflow to nuclear SMBHs occur from kiloparsec-scales, and thus controlling the mass accretion rate of the accreting BH. Another interesting paradigm to explore the possible effects of $\Lambda$, could be in the area of AGN feedback, where outflows, jets and radiation from the nuclear region of the active galaxies harboring SMBHs interact with the ISM and beyond, with length-scales ranging over several kiloparsecs to even megaparsecs. The feedback might considerably effect the star formation rates as well as can curtail accretion onto host SMBH, and thus may influence the formation and evolution of galaxies [49,50]. There would be significant effects of $\Lambda$ at these 
length-scales in the expanding Universe. 

To have a glimpse on how repulsive $\Lambda$ could influence the realistic astrophysical phenomena as well as to examine the effectiveness of the potential derived here, as a toy model, we investigated the impact of $\Lambda$ on Bondi accretion rate, which approximately governs the accretion dynamics plausibly in low luminous AGNs or in LERGs. Using SDS analogous potential, it is found that the current accepted value of cosmological constant impact the sonic radius as well as the Bondi accretion rate moderately, for an ambient temperature $T_{\infty} < 10^4 K$ and for smaller values in $\gamma$ ($\lsim \, 1.5$). Thus, it appears that the said value of repulsive cosmological constant might influence the Bondi accretion phenomenon, with measurable effect in the local galactic-scales. This might have interesting consequences, especially on the dynamics of advective accretion flow; the detailed study of which would be pursued in the near future. Note that, [19] also analyzed the influence of $\Lambda$ on Bondi accretion rate, however with significantly large values of $\Lambda$. Interestingly the authors of [19] adopted unrealistic high value of $\Lambda$, however because for small $\Lambda$ their adopted technique becomes too difficult under pure GR framework.     

The next goal is to test the SDS analogous potential in regard to the realistic astrophysical processes in local galactic-scales in order to explore the influence of $\Lambda$ on kiloparsecs-scale structure and beyond, in the local-galaxies. 

\section*{Acknowledgments}
The authors are thankful to an anonymous reviewer for insightful comments and suggestions that help us to improve the manuscript.

\appendix
\section{}

Here we furnish the Cartesian transformation of the acceleration terms of the test particle motion using potential $V_{\rm DS}$ in SDS back ground, which are useful to depict the elliptical like trajectory of the particle orbit. The acceleration of the test particle in $x^i$ direction, where $x^i = x, y, z$ is given by 

\begin{eqnarray}
{\ddot{x}}^i = \frac{x^i}{r}  fs(r) + fs1(r) \, {\dot x}^i \, \sum_i x^i {\dot x}^i \, ,
\label{A1}
\end{eqnarray}

where,  

\begin{eqnarray}
fs(r) = \left(-\frac{r_s}{r^2} + \frac{\Lambda r}{3} \right) \left[\frac{f(r)}{r}\right]^2 - \frac{3}{r^4} \, \sum_i \left(\sum_{jk} \epsilon_{ijk} x^j {\dot x}^k    \right)
\label{A2}
\end{eqnarray}

and 

\begin{eqnarray}
fs1(r) = 2 \, \frac{r_s-\frac{\Lambda r^3}{3} }{r^2 f(r)} \, ,  
\label{A3}
\end{eqnarray}

respectively. $f(r) = r - 2r_s- \frac{\Lambda r^3}{3}$ and $r^2 =  \sum_i x^i x^i$. 

%**********************************************************************

\end{document}